\documentclass[graybox]{svmult}

\usepackage{type1cm}        
\usepackage{makeidx}         
\usepackage{graphicx,graphics,psfrag}        
\usepackage{multicol}        
\usepackage[bottom]{footmisc}

\usepackage{newtxtext}       %
\usepackage[varvw]{newtxmath}       

\usepackage{bbold, epsfig, euscript,enumitem}

\newcommand{\divS}{{\rm div}_\Sigma \,}

\renewcommand{\div}{{\rm div \,}}

\newcommand{\esssup}[1]{\mathop{\rm ess\ sup}}
\newcommand{\essinf}[1]{\mathop{\rm ess\ inf}}
\newcommand{\N}{{\rm I\kern - 2.5pt N}}
\newcommand{\Z}{{\rm Z\kern - 5.5pt Z}}
\newcommand{\Q}{{\rm I\kern - 5.25pt Q}}
\newcommand{\C}{{\rm I\kern - 6.25pt C}}
\newcommand{\R}{{\rm I\kern - 2.5pt R}}
\newcommand{\bbf}{\mathbf{b}}

\newcommand{\Dbf}{\mathbf{D}}
\newcommand{\fbf}{\mathbf{f}}

\newcommand{\nbf}{\mathbf{n}}

\newcommand{\qbf}{\mathbf{q}}

\newcommand{\Sbf}{\mathbf{S}}
\newcommand{\ubf}{\mathbf{u}}
\newcommand{\vbf}{\mathbf{v}}

\newcommand{\Phibf}{\mathbf{\Phi}}
\newcommand{\na}{\nabla}
\newcommand{\pa}{\partial}
\newcommand{\stress}{\mathbf{S}}

\makeindex             

\begin{document}

\title*{Multi-Velocity Sharp-Interface Continuum Thermodynamics of Fluid Systems with Adsorption}
\titlerunning{Multi-velocity sharp-interface continuum thermodynamics}
\author{Dieter Bothe\vspace{0.15in}\\
\textsl{In memoriam Wolfgang Dreyer}}
\authorrunning{Dieter Bothe}
\institute{Dieter Bothe \at Department of Mathematics and Profile Topic \emph{Thermofluids \&  Interfacial Phenomena}, Technical University of Darmstadt, Peter-Gr\"{u}nberg-Stra{\ss}e~10,
D-64287 Darmstadt, Germany.
\email{bothe@mma.tu-darmstadt.de}}
%
%
\maketitle

\abstract*{We revisit the sharp-interface continuum thermodynamics of two-phase multicomponent fluid systems, accounting for partial mass and partial momentum balances both in the bulk phases and on the interface. This allows to describe the transfer of species between the individual bulk phases and the interface, i.e.\ ad- and desorption processes. In fact, the transfer of any constituent between the two bulk-phases is considered as a series of ad- and desorption processes. In this framework, all species transfer
processes are coupled via the interfacial thermodynamics.
As a consequence, the influence of surface active species on the transfer of other constituents can be captured in detail. 
The derivation of this model class relies on an axiomatic form of the entropy principle which, at the same time, allows for an efficient closure process.
This form of the entropy principle has been introduced for one-phase fluid systems  in \cite{BD2015} as the result of intense joint work of the late Wolfgang Dreyer and the present author.}

\abstract{We revisit the sharp-interface continuum thermodynamics of two-phase multicomponent fluid systems, accounting for partial mass and momentum balances both in the bulk phases and on the interface. This allows to describe the transfer of species between the bulk phases and the interface, i.e.\ ad- and desorption processes. In fact, the transfer of any constituent between the two bulk-phases is considered as a series of ad- and desorption processes. 
In this framework, all species transfer
processes are coupled via the interfacial thermodynamics.
As a consequence, the influence of surface active species on the transfer of other constituents can be captured in detail. 
The derivation of this model class relies on an axiomatic form of the entropy principle which, at the same time, allows for an efficient closure process.
This form of the entropy principle has been introduced for one-phase fluid systems  in \cite{BD2015} as the result of intense joint work of the late Wolfgang Dreyer and the present author.}

\section{Introduction}
Multicomponent two-phase fluid systems are ubiquitous in nature, science and engineering.
A prototypical case is that of an air bubble in water.
For this system, water, nitrogen, oxygen and carbon dioxide, among many more, are the involved constituents.
A generic two-phase fluid system thus is composed of two multicomponent mixtures which form the two bulk phases, being in contact at their common interface.
We assume that the chemical species are homogeneously mixed on the molecular scale, where the composition of the mixture depends on the position $\bf x$ and on the time instant $t$. In a continuum physical description, balance equations for the partial masses, i.e.\ individual balance equations for the species mass densities $\rho_i$, are required, where the species are indexed with $i=1,\ldots,N$ in case of $N$ different chemical components.

We consider two-phase fluid systems inside a fixed domain $\Omega \subset \R^3$ in which two bulk phases, occupying the subdomains $\Omega^+ (t)$ and $\Omega^- (t)$, are separated by an interface $\Sigma (t)$.
This interface represents an extremely thin transition zone between the two bulk phases, in which the composition will change abruptly. 
As time evolves, the interface may move and deform.
We employ the sharp-interface modeling framework, in which $\Sigma (t)$ is represented by a family of hypersurfaces, i.e.\ this transition layer is modeled as having zero thickness. This is a very sensible approximation under conditions (temperature, pressure) away from the so-called critical point at which this transition zone becomes more and more thicker before it finally disappears as the fluid system changes into a single-phase supercritical state. For water, the critical point is at about 647~K and 22 MPa ($\approx 220$ bar); see \cite{Water}.
Under so-called normal conditions (i.e., at a temperature of $20^\circ$ C $=293.15$ K and a pressure of $p=1$ bar $=101\, 325$ Pa), the water-air interface, for instance, has a thickness of about 0.6-0.8 nm, defined as the thickness of the zone of anisotropic structure.
While this surprisingly small thickness was already known from molecular dynamics simulations for some time, it has only been measured very recently \cite{fellows2024thick}. This underlines the validity of the sharp-interface approximation.
In this idealized scenario, the partial mass densities as well as all other intensive quantities, display a jump discontinuity at $\Sigma$. 
For example, at the (planar) air-water interface and at normal conditions,
the oxygen mass density jumps from $\rho_{O_2}^L = 278.6\; {\rm g\, m}^{-3}$ in the liquid phase to $\rho_{O_2}^G = 9.1\; {\rm g\, m}^{-3}$ in the gas phase.
Hence the oxygen density jumps down to only $3.3 \%$.
Therefore, an integral part of any sharp-interface model is formed by so-called jump conditions, relating the two one-sided limits of, e.g., the species mass densities.

In many real world fluid systems, surface active substances (so-called \emph{surfactants}) are present, which adsorb at the fluid interface, changing the interfacial free energy.
If $A_i$ denotes such a surfactant species, the interface will locally carry a certain mass of $A_i$ per interfacial area, hence an area-specific mass density
$\rho_i^\Sigma$ is to be used to describe the presence of $A_i$ in the adsorbed state.
This is mandatory in order to add a proper interfacial thermodynamical description.
The process of transfer of a species from one of the bulk phases to the interface is called \emph{adsorption}.
In order to distinguish between a species either as present in the bulk phases or in the adsorbed state on the interface, we will use the superscripts $+, -$ and $\Sigma$, such that $A_i^\Sigma$ denotes the $i$-th constituent in its adsorbed form. If the adsorbed form $A_i^\Sigma$ is present, this species will typically also be transferred into one or both of the adjacent bulk phases, a process called \emph{desorption}.

Two-phase fluid systems which are out of equilibrium exchange, in particular, mass, momentum and energy.
In multicomponent two-phase fluid systems away from 'chemical equilibrium' (to be made precise below), matter will be exchanged across the interface,
i.e.\ a transfer of chemical constituents takes place. This process will be called \emph{mass transfer} throughout.
For instance, if air is brought into contact with pure water, then oxygen and nitrogen (as well as all other gaseous species inside the air) will be transferred into the water, i.e.\ the bubble will start to dissolve. In a more realistic scenario, the water will already carry some dissolved gaseous components. Once the gas and the liquid phase come into contact, these dissolved components will be transferred from the surrounding liquid into the bubble unless they are already present there at sufficiently high concentrations.
Hence different species are transferred simultaneously and, possibly, in different directions -- so-called conjugate mass transfer.
This has been quantitatively demonstrated in \cite{hosoda2014mass}, where the dissolution of a ${\rm CO}_2$-bubble in water has been investigated. The experiment showed that the bubble will not disappear completely and by gas chromatography it was shown that the residual bubble was composed of air, coming out of the liquid phase, where it was dissolved in the water.

Adsorption of surfactant strongly influences the transfer rates of, e.g., gaseous components into a liquid phase.
It is commonly accepted that there are two different ways, how this influence is mediated.
First, since surfactant is inhomogeneously distributed along the interface, the interfacial tension displays non-zero
surface gradients. This induces so-called Marangoni stresses which act as tangential forces at the interface, thus changing the flow field locally, resulting for instance in a reduced terminal rise velocity of buoyant bubbles.
Second, even in the absence of fluid flow, the partial coverage of the interface with surfactant molecules constitutes a barrier against the passage of other molecules across the interface. This leads to an additional \emph{hindrance effect}, also referred to as mass transfer resistance, sometimes also denoted a steric effect. Experimental results on this hindrance effect can be found, e.g., in
\cite{Sardeing2006, Hebrard2009, Aoki2015, Hori2019, Hori2020}.
Let us note that the analogous phenomenon on the rate of evaporation, being strongly influenced by the presence of surfactants, has been studied much earlier by Langmuir; see \cite{Langmuir}, where a so-called energy barrier model was introduced to explain this phenomenon, at least qualitatively.
A survey on the surfactant-induced hindrance effect onto evaporation and mass transfer can be found in \cite{Barnes1986}.

The influence of surfactants on local mass transfer rates due to Marangoni effects has been studied recently by means of numerical simulations on different levels of detail, see \cite{kentheswaran2023impact, bubbly-flows-ijmf} and the references given there.
In the underlying mathematical models, the presence of surfactant changes the local surface tension.
The resulting Marangoni stresses strongly impact the local hydrodynamics and, hence, the mass transfer rates mainly due to a modified interplay
of diffusive and convective characteristic time scales.
However, the direct hindrance effect of surfactant on mass transfer processes is not included in these models.
If it is accounted for at all, then via global mass transfer correlations; see, e.g., \cite{Sardeing2006}.
A possible reason behind is a fundamental difficulty to include such hindrance effects as a \emph{local} mechanism into continuum physical models:
in the standard models, the mass transfer of, for example, a gaseous species is driven by the difference of the one-sided limits of the chemical potential at the interface. 
These limit values depend on bulk quantities such as, with a common choice of primitive variables, the local temperature, pressure and composition at the respective position. If surfactants are included in this standard model, their effect is a change of the local surface tension. While this affects the local hydrodynamics
via the momentum jump condition, a model without \emph{interfacial} (molar) mass for all relevant species does not provide a channel for a direct interaction between the local coverage by surfactant and the local rate of mass transfer of another constituent.

\begin{figure}
\sidecaption
\includegraphics[width=2.0in]{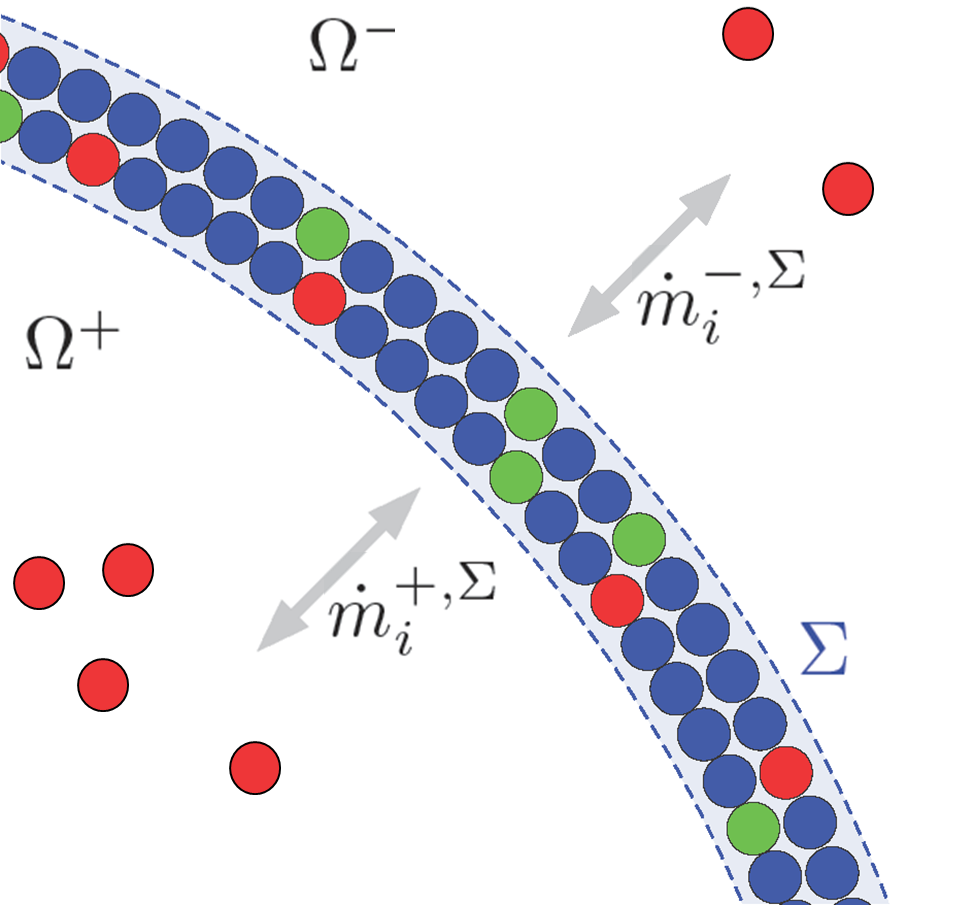}
\caption{Mass transfer between two bulk phases as a series of two one-sided bulk-interface transfer processes.}\label{transfer-series}
\end{figure}

To implement such a channel of interaction between adsorbed surfactant and a transfer species,
the model needs to be enriched. Note that, if a species is transferred from one side of the interface to the other one,
it has to pass the transition layer in between, hence will be present within this layer for some positive residence time.
See Figure~\ref{transfer-series} for an illustration of this conceptual consideration.
%
%
%
%
This implies that, within the framework of a sharp-interface model, the average number of particles (atoms or molecules) of this transfer species within
a certain surface area within $\Sigma$ is positive. In other words, if $A_k$ is this transfer species, than $\rho_k^\Sigma >0$.
The point here is not about possible inertia effects of this contribution to the mass on the interface, but on the implications for the thermodynamical state of the interface as a multicomponent interfacial mixture. 
Consequently, to allow for proper coupling of mass transfer and sorption phenomena, \emph{all constituents} should be included in the model with 
partial mass densities in the bulk phases and also on the interface. Of course, this still allows for a specific constituent which is only present in one bulk phase or even just on the interface. This idea has been introduced and investigated in \cite{Bo-interface-mass} by the present author.
Let us note in passing that surface concentrations in this model class are necessarily non-negative, while so-called excess mass densities
for the dividing interface due to Gibbs can be negative; see \cite{Guggenheim}.
This non-negativity is a qualitative property of solutions of our model, which implies certain structural properties of the transfer coefficients; cf.\ \cite{BD-MCD} for the implications of positivity requirements on (cross-)diffusion coefficients.
The price to pay for having this useful property is that the spatial region which is lumped into the sharp interface is still considered as bulk volume with the bulk densities. This introduces a negligible error, unless the volume of the transition zone becomes a significant fraction of the total volume.

This conceptual picture, as illustrated by Figure~\ref{transfer-series}, also implies that mass transfer must be considered
as the succession of two bulk-interface transfer processes, taking place one after the other.
Furthermore, both of these one-sided mass transfer processes are bi-directional, which opens a more flexible and realistic way of modeling.
In this picture, the interfacial thermodynamical state automatically influences both one-sided bi-directional mass transfer rates, hence also
the resulting total rate of the series of transfer steps.

Mathematical models for multicomponent two-phase fluid systems that account for the interfacial mass densities of all constituents have only been  rarely investigated in the literature.
Building on prior sharp-interface modeling of irreversible processes given in \cite{Waldmann} and \cite{BAM},
in the seminal paper \cite{Bedeaux}, Dick Bedeaux derived such a sharp-interface model with entropy inequality
in the spirit of T.I.P.\ (Theory of Irreversible Processes, cf.\ De Groot, Mazur \cite{dGM}).
A few years later, the monograph \cite{Slattery-Interfaces} by Slattery appeared, which contains a similar
sharp-interface model with interfacial mass densities, deriving and evaluating bulk and interfacial entropy production rates in the spirit of Truesdell and Noll; cf.\ \cite{TN2003}. 
In these works, and later in \cite{Sagis, Dreyer-Guhlke-M, Bo-interface-mass}, an interfacial entropy production rate is given.
Most of the given entropy production rates are not equivalent and those which are correct are often arranged in a disadvantageous way such that a resulting closure, if given at all, would not result in a physically sound model.

In the present paper, we derive a general sharp-interface model for multicomponent
two-phase fluid systems in which not only partial mass densities in the bulk phases and
on the interface are included for all constituents, but also \emph{partial} momenta of all species in the bulk phases and
on the interface are individually balanced.
Hence, we develop the \emph{multi-velocity} sharp-interface model class for chemically reacting fluid systems composed of $N$ chemical constituents.
In the notion of Hutter \cite{Hutter-book}, this yields a \emph{class-II} model for fluid systems with interfacial mass, where every constituent
has its individual velocity, with independent bulk and interface fields.
Our main motivation to include partial momentum balances with their individual, species-specific velocities is the quest for a model class with a minimum number of necessary cross-couplings in order to obtain a practically useful PDE model. This requires to properly relate all terms appearing in the entropy production with corresponding dissipative mechanisms, which in turn relies on a clarification of the structure of the entropy production. Here it turns out that the class-II, i.e.\ the multi-velocity model reveals such relations that are hidden if a more standard class-I model (with the common, barycentric velocity) is employed. 
The latter case was investigated in  \cite{Bo-interface-mass}, which treated the multicomponent two-phase fluid systems with interfacial mass, putting main emphasis on the surfactant hindrance effect on mass transfer.
There, in order to clarify the structure of the entropy production, the class-II balance equations and the entropy production rates for bulk phases and the sharp interface has been collected in an appendix. Here, we resume the investigations from \cite{Bo-interface-mass}, presenting a self-contained derivation of the model and adding in particular a full set of constitutive equations. The multi-velocity approach also allows for a refined and more detailed modeling of interfacial transport processes. In principle, a further extension of this model to account for individual temperatures is also possible; see \cite{ruggeri2009multitemp} concerning such multi-temperature modeling in case of single-phase fluid mixtures.
 
Many more details about the
general strategy of deriving the class-II continuum thermodynamics can be found in the joint work \cite{BD2015} of Wolfgang Dreyer and the present author, where many new ideas on the continuum thermodynamics of single-phase multicomponent chemically reacting fluid mixtures have been developed. The present work also profited from Wolfgang's 'jump paper' \cite{6}.
\section{Sharp-Interface mathematical modeling}
We assume the two-phase multicomponent fluid system to fill the considered domain $\Omega \subset \R^3$
in such a way that two bulk phases
$\Omega^\pm (t)$ are separated by a sharp interface, i.e.\ by a surface $\Sigma (t)$ in $\R^3$:
\[
\Omega = \Omega^+ (t) \cup \Omega^- (t) \cup \Sigma (t),
\]
a disjoint decomposition of $\Omega$. For technical simplicity, we assume the $\Sigma (t)$ to be closed surfaces without
contact to the domain boundary $\partial \Omega$, thus avoiding the appearance of contact lines.
Let us note in passing that this setup assumes the total domain to be fixed, hence the total volume is conserved if $\Omega$ is a bounded set.
This is a standard assumption in essentially all mathematical models, as it allows to work with a set of PDEs inside a fixed volume.
A more realistic approach would allow (some part of) the boundary to have some flexibility and, e.g., to introduce a pressure control; cf.\ \cite{bothe2016thermodynamically}.

The family of moving hypersurfaces $\{\Sigma (t) \}_{t\in I}$, with $I=(a,b)\subset \R$ an open time interval, is assumed to form a
$\mathcal{C}^1$-hypersurface in space-time $\R^4$ such that the instantaneous surfaces $\Sigma (t)$ are $\mathcal{C}^2$-surfaces
in $\R^3$ so that curvature is well-defined on $\Sigma (t)$. In particular, the total curvature (twice the mean curvature)
$\kappa_\Sigma =\divS (- {\bf n}_\Sigma )$ is a well-defined continuous quantity. In addition, we assume that the normal field
${\bf n}_\Sigma (t,\cdot)$ of $\Sigma (t)\subset \R^3$ is continuously differentiable, jointly in both variables $(t,x)$.
Under these assumptions, we call $\{\Sigma (t) \}_{t\in I}$ a $\mathcal{C}^{1,2}$-family of moving hypersurfaces.
We also denote by $\mathcal M ={\rm gr}\,( \Sigma )$ the graph of the family $\{\Sigma(t)\}_{t \in I}$, defined as
\begin{equation}
\mathcal M := \{ (t,{\bf x})\in \R^4 : t\in I, \, {\bf x}\in \Sigma (t)\} = \bigcup_{t\in I} \big( \{t\}\times \Sigma (t) \big).
\end{equation}
These definitions are also employed in \cite{Kimura.2008}, \cite{PrSi15}, \cite{Bo-interface-mass} and in a similar form in \cite{Giga.2006}.

For a $\mathcal{C}^{1,2}$-family $\{\Sigma(t)\}_{t \in I}$ of moving hypersurfaces, $V_\Sigma$ denotes
the \emph{speed of normal displacement} of $\Sigma(\cdot)$ and is defined via the relation
\begin{equation}\label{VSigma0}
V_\Sigma (t,{\bf x}) = \langle \gamma' (t) , {\bf n}_\Sigma (t, \gamma (t)) \rangle
\end{equation}
for any $\mathcal{C}^1$-curve $\gamma$ with $\gamma (t)={\bf x}$ and ${\rm gr} \,(\gamma)\subset \mathcal M$,
where $\langle \cdot , \cdot \rangle$ denotes the inner product in $\R^3$.
It is not difficult to show that the definition via \eqref{VSigma0} is equivalent to
\begin{equation}\label{VSigma}
V_\Sigma (t,{\bf x}) = -{\bf n}^t_\mathcal{M}(t,{\bf x}) / || {\bf n}^{\bf x}_\mathcal{M}(t,{\bf x}) ||,
\end{equation}
where ${\bf n}_\mathcal{M} = ( {\bf n}^t_\mathcal{M}, {\bf n}^{\bf x}_\mathcal{M})$ is the normal field of $\mathcal M$, oriented in such a way that
${\bf n}_\Sigma (t,{\bf x})={\bf n}^{\bf x}_\mathcal{M}(t,{\bf x}) / || {\bf n}^{\bf x}_\mathcal{M}(t,{\bf x}) ||$.
The characterization via \eqref{VSigma} shows that $V_\Sigma$ is a purely kinematic quantity, determined only by the family $\{\Sigma(t)\}_{t \in I}$ of moving interfaces. Hence, the value of $V_\Sigma$ from \eqref{VSigma0}
does not depend on the choice of the specific curve; cf.\ Chapter~2.5 in \cite{PrSi15}.
The computation of $V_\Sigma$ is especially simple
if $\{\Sigma(t)\}_{t \in I}$ is given by a level set description, i.e.\
\begin{equation}\label{levelset}
\Sigma(t)=\{{\bf x} \in \R^3: \phi(t,{\bf x})=0\}
\end{equation}
with $\phi \in \mathcal{C}^{1,2}(\mathcal N)$ for some open neighborhood $\mathcal N \subset \R \times \R^3$ of
$\mathcal M$ such that $\nabla_{\bf x} \phi \not= 0$ on $\mathcal{M}$. In this case, ${\bf n}_\mathcal{M} = \nabla_{(t,{\bf x})} \phi /||\nabla_{(t,{\bf x})} \phi || = (\partial_t \phi, \nabla_{\bf x} \phi )/||\nabla_{(t,{\bf x})} \phi ||$, hence
\begin{equation}\label{E11}
V_\Sigma(t,{\bf x})=- \, \frac{\partial_t \phi(t,{\bf x})}{\|\nabla_{\bf x} \phi(t,{\bf x})\|}\quad\text{ for } t \in J, \, {\bf x} \in \Sigma(t).
\end{equation}
In the literature, $V_\Sigma$ is often called normal velocity of $\Sigma (\cdot)$, but we prefer to call it the speed
of normal displacement since $V_\Sigma$ is a scalar quantity.
In fact, the speed of normal displacement gives rise to a natural interface velocity field,
given as ${\bf w}_\Sigma:=V_\Sigma {\bf n}_\Sigma$, which therefore is intrinsic to any $\mathcal{C}^{1,2}$-family $\{\Sigma(t)\}_{t \in I}$ of moving hypersurfaces.
According to Corollary~1 together with Theorem~1 in \cite{Bo-2PH-ODE}, the associated initial value problem
\begin{equation}\label{2PH-IVP}
\dot {\bf x} (t) = {\bf w}_\Sigma (t,{\bf x}(t)) \quad \mbox{ for } t\in I,\;\; {\bf x}(t_0)={\bf x}_0
\end{equation}
is uniquely solvable (at least locally in time) for every $t_0\in I$ and ${\bf x}_0\in \Sigma (t_0)$.
The unique (local) solution ${\bf x}(\cdot \hspace{1pt}; t_0,{\bf x}_0)$ of \eqref{2PH-IVP}, starting in $(t_0,{\bf x}_0)$,
follows the moving surface along a path which always runs normal to $\Sigma(t)$.
Consequently, the intrinsic interface velocity ${\bf w}_\Sigma$ (hence $V_\Sigma$)
induces a Lagrangian-type derivative, the so-called
Thomas derivative $\partial^\Sigma_t$ (cf.\ \cite{Thomas}), defined as
\begin{equation}\label{Thomas-derivative}
\partial^\Sigma_t \phi^\Sigma (t_0,{\bf x}_0) := \frac{d}{dt} \phi^\Sigma (t,{\bf x}(t;t_0,{\bf x}_0))_{|t=t_0},
\end{equation}
where $\phi^\Sigma$ denotes a quantity which is defined on $\mathcal{M}$.
According to Lemma~2 together with Theorem~1 in \cite{Bo-2PH-ODE}, unique (local) solvability of the associated
initial value problems also holds true for any velocity field of the type\footnote{We use $\Sigma$ as a subscript for intrinsic, geometrically defined quantities, while
we use $\Sigma$ as superscript for all other interface quantities.}
\begin{equation}\label{full-velocity}
{\bf v}^\Sigma = V_\Sigma \, {\bf n}_\Sigma + {\bf v}^\Sigma_{||},
\end{equation}
where the additional tangential part ${\bf v}^\Sigma_{||}$ is continuous on $\mathcal{M}$ and
the maps ${\bf v}^\Sigma_{||} (t,\cdot )$ are locally Lipschitz continuous on $\Sigma (t)$ for every $t\in I$.
This gives rise to a Lagrangian derivative $\frac{D^\Sigma}{Dt}$
associated with ${\bf v}^\Sigma$, defined in analogy to the Thomas derivative above, for which one can show that
\begin{equation}\label{Lagrange-Thomas}
\frac{D^\Sigma \phi^\Sigma}{Dt} = \partial^\Sigma_t \phi^\Sigma + {\bf v}^\Sigma_{||} \cdot \nabla_\Sigma \phi^\Sigma
\end{equation}
with $\nabla_\Sigma$ denoting the surface gradient (w.r.\ to $\Sigma (t)$, where $t$ is understood); cf.\ \cite{Bo-2PH-ODE}.
Note also that ${\bf v}^\Sigma_{||} \cdot \nabla_\Sigma \phi^\Sigma = {\bf v}^\Sigma \cdot \nabla_\Sigma \phi^\Sigma$,
since $\nabla_\Sigma \phi^\Sigma$ is tangential to $\Sigma$.

The Thomas derivative allows to formulate a general surface transport theorem.
For this purpose, let $V\subset \Omega$ be a fixed control volume with outer normal field
${\bf n}$ and let $\Sigma_V (t):=\Sigma (t) \cap V$.
Then the identity
\begin{equation}\label{2PH-trans-thm0}
\frac{d}{dt} \int_{\Sigma_V (t)} \!\! \phi^\Sigma \,do = \int_{\Sigma_V (t)}\!\!  \big( \partial_t^\Sigma \phi^\Sigma
- \phi^\Sigma \kappa_\Sigma V_\Sigma \big)\,do
 - \int_{\partial \Sigma_V (t)}\!\!   \frac{\phi^\Sigma V_\Sigma \, {\bf n}\cdot {\bf n}_\Sigma}{\sqrt{1-({\bf n}\cdot \mathbf{n}_{\Sigma})^2}}\,ds
\end{equation}
holds, see \cite{Romano, Gurtin, Alke}.
For a particular control volume $V$ such that ${\bf n}$ is tangential to $\Sigma_V (t)$ on $\partial \Sigma_V (t)$,
i.e.\ ${\bf n} \cdot {\bf n}_\Sigma =0$ on $\partial \Sigma_V (t)$, the transport identity \eqref{2PH-trans-thm0} simplifies to
\begin{equation}\label{2PH-trans-thm1}
\frac{d}{dt} \int_{\Sigma_V (t)} \!\! \phi^\Sigma \,do = \int_{\Sigma_V (t)}\!\!  \big( \partial_t^\Sigma \phi^\Sigma
- \phi^\Sigma \kappa_\Sigma V_\Sigma \big)\,do.
\end{equation}
Below, any control volume which intersects $\Sigma (t)$ is assumed to be of this type,
which is sufficient for deriving the local balance equations.
From here on, the time dependence of $\Sigma_V (t)$ and $\partial \Sigma_V (t)$ in integration domains
is suppressed for better readability.

\begin{figure}
\sidecaption
\includegraphics[width=2.6in]{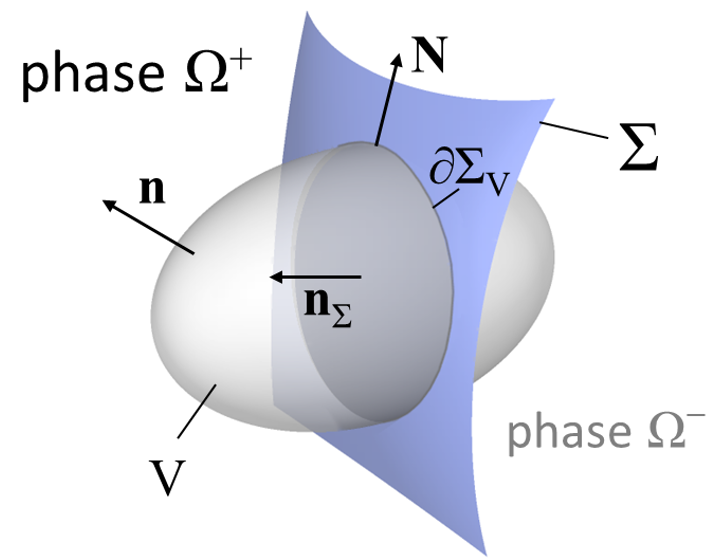}
\caption{Configuration of the phases and basic notations.}\label{CV-fig}
\end{figure}

To obtain a generic balance equation, we start with {\it integral balances}, since they are valid across the interface,
while a direct formulation of interfacial transmission and jump conditions is not obvious.
Let $\Phi$ denote an extensive, i.e.\ a mass additive quantity such as mass, momentum, energy,
with volumetric density $\phi$ (i.e.\ $\phi^+$ and $\phi^-$ in the phase
$\Omega^+$ and $\Omega^-$, respectively) and interfacial, area specific density $\phi^\Sigma$.
Given a fixed control volume $V$, 
let $\mathbf{N}$ denote the outer
unit normal to $\partial \Sigma_V$ which is at the same time tangential to $\Sigma$; cf.\ Figure~\ref{CV-fig}.
With these notations, the generic integral balance for the amount of an extensive quantity $\Phi$ inside a \emph{fixed} control volume $V$ is given by
\begin{eqnarray}
\label{integral-master}
\frac{d}{dt} \big[ \int_{V} \phi\,dx + \int_{\Sigma_V} \phi^\Sigma \,do \big]
& = &
-\int_{\partial V} \mathbf{j}_{\rm tot}\cdot\mathbf{n}\,do +\int_{V} f\,dx\\[1ex]\nonumber
& & - \int_{\partial \Sigma_V} \mathbf{j}^{\Sigma}_{\rm tot}\cdot\mathbf{N}\,ds +\int_{\Sigma_V} f^{\Sigma}\,do,
\end{eqnarray}
where $\mathbf{j}_{\rm tot}$ is the flux of $\Phi$ inside the bulk phases, $\mathbf{j}^{\Sigma}_{\rm tot}$
the corresponding flux inside the interface, $f$ the source terms acting in the
bulk and, finally, $f^\Sigma$ the interfacial source term.
It is important to note that both $\mathbf{j}_{\rm tot}$ and $\mathbf{j}^\Sigma_{\rm tot}$ are total fluxes
with respect to the reference frame in which the fixed control volume is given.
Below, they will be split into a convective plus a diffusive contribution.
We require the interfacial fluxes to be tangential
to the interface. This is no restriction, since only $\mathbf{j}^{\Sigma}_{\rm tot}\cdot\mathbf{N}$ enters the balance equation.
In order to obtain the local form of the generic balance equation, some mathematical tools are required.
We briefly recall the relevant identities, while more details can be found in \cite{Slattery-etal-Interfaces};
for proofs see \cite{Kimura.2008, PrSi15} or \cite{Garcke-Handbook}.
The two-phase transport theorem in the version for fixed control volumes $V$ states that
\begin{equation}\label{2PH-trans-thm-Euler}
\frac{d}{dt} \int_{V} \phi \,dx = \int_{V\setminus \Sigma} \partial_t \phi \, dx
- \int_{\Sigma_V} [\![ \phi ]\!] V_\Sigma \,do,
\end{equation}
where
\begin{equation}
\label{jump-notation}
[\![ \phi ]\!] (t,{\bf x}) := \lim_{h\to 0+} \big( \phi (t,{\bf x}+h \mathbf{n}_{\Sigma})
- \phi (t,{\bf x}-h \mathbf{n}_{\Sigma}) \big)
\end{equation}
denotes the jump of a quantity $\phi$ across the interface in a form which makes the jump conditions orientation-invariant.
For later use, note that 
\begin{equation}
\label{jump-resolved}
[\![ \phi ]\!] \, {\bf n}_\Sigma = - \, \big( \phi^+ {\bf n}^+ + \phi^+ {\bf n}^+ \big),
\end{equation}
where ${\bf n}^\pm$ denotes the outer normal to $\Omega^\pm$ at $\Sigma$.
The two-phase divergence theorem provides the identity
\begin{equation}\label{2PH-div-thm}
\int_{\partial V} {\bf f}\cdot {\bf n} \,do = \int_{V\setminus \Sigma} \div {\bf f}\, dx
+ \int_{\Sigma_V} [\![ {\bf f}\cdot {\bf n}_\Sigma ]\!] \,do.
\end{equation}
We intentionally write $V\setminus \Sigma$ as the domain of integration to avoid confusion with distributional derivatives which would include a Dirac delta contribution supported on the interface; the latter would in fact correspond to the jump bracket term in \eqref{2PH-div-thm}.
Finally, the surface divergence theorem states that
\begin{equation}\label{int-divM-f}
\int_{\partial \Sigma_V} \mathbf{f}^{\Sigma} \cdot {\bf N} \,ds
= \int_{\Sigma_V} {\rm div}_\Sigma \, \mathbf{f}^{\Sigma} \, do
\end{equation}
for vector fields $\mathbf{f}^{\Sigma}$ being tangential to $\Sigma$.

Applying the two-phase transport theorem to the first term in \eqref{integral-master},
the extended surface transport theorem in the form \eqref{2PH-trans-thm1}
to the second term, the two-phase divergence theorem to the third term and the surface divergence theorem
to the fifth term yields
\begin{eqnarray}
\label{integral-master3}
\nonumber
\int_{V\setminus \Sigma} ( \partial_t \phi + \div {\bf j}_{\rm tot} ) \, dx
+ \int_{\Sigma_V} (-[\![ \phi ]\!] V_\Sigma + [\![ \mathbf{j}_{\rm tot} \cdot \mathbf{n}_{\Sigma} ]\!] )\, do\\[1ex]
+ \int_{\Sigma_V} \big( \pa_t^\Sigma \phi^\Sigma + \divS (\phi^\Sigma V_\Sigma {\bf n}_\Sigma + {\bf j}^\Sigma_{\rm tot}) \big)\, do  = \int_{V} f \, dx  + \int_{\Sigma_V} f^{\Sigma}\,do.\nonumber
\end{eqnarray}
Collecting bulk and interface terms, we obtain the differential form of the generic balance equation in an Eulerian form
by means of the usual localization procedure.
The result reads as
\begin{eqnarray}
\label{master-bal-E1}
\partial_t \phi + \div {\bf j}_{\rm tot} =f & \mbox{ in } \Omega\setminus \Sigma,\\[1ex]
\pa_t^\Sigma \phi^\Sigma + \divS (\phi^\Sigma V_\Sigma {\bf n}_\Sigma + {\bf j}^\Sigma_{\rm tot})
 -[\![ \phi ]\!] V_\Sigma + [\![ \mathbf{j}_{\rm tot} \cdot \mathbf{n}_{\Sigma} ]\!] =f^\Sigma & \mbox{ on } \Sigma.
\end{eqnarray}
Recall that both $\mathbf{j}_{\rm tot}$ and $\mathbf{j}^\Sigma_{\rm tot}$ are total fluxes with respect to the
reference frame in which the chosen control volumes are fixed, and $\mathbf{j}^\Sigma_{\rm tot}$ is tangential to $\Sigma (t)$.
If these are decomposed into a convective and a diffusive (also called molecular) flux according to
\begin{equation}
\label{flux-decomposition}
\mathbf{j}_{\rm tot}= \phi \, {\bf v} +  \mathbf{j}, \qquad
\mathbf{j}^\Sigma_{\rm tot}=  \phi^\Sigma  \, {\bf v}^\Sigma_{||} +  \mathbf{j}^\Sigma,
\end{equation}
the generic local balance equations for an extensive quantity $\Phi$ with bulk density $\phi$ and interfacial
density $\phi^\Sigma$ read as
\begin{eqnarray}
\label{master-bal-E2}
\partial_t \phi + \div \big( \phi \, {\bf v} +  \mathbf{j} \big) =f & \mbox{ in } \Omega\setminus \Sigma,\\[1ex]
\pa_t^\Sigma \phi^\Sigma + \divS (\phi^\Sigma  \, {\bf v}^\Sigma +  \mathbf{j}^\Sigma)
+ [\![ \phi ({\bf v} - {\bf v}^\Sigma )]\!] \cdot \mathbf{n}_{\Sigma}  + [\![ \mathbf{j} \cdot \mathbf{n}_{\Sigma}  ]\!] =f^\Sigma & \mbox{ on } \Sigma.
\end{eqnarray}
Concerning the jump bracket notation, we treat interfacial quantities as having one-sided bulk limits of the same value,
e.g.\ $[\![ {\bf j}\cdot \mathbf{n}_{\Sigma} ]\!]$ above is the same as $[\![ {\bf j} ]\!]\cdot \mathbf{n}_{\Sigma}$.
Hence, the generic interfacial balance can be written somewhat more compactly as
\begin{equation}
\label{master-bal-int}
\pa_t^\Sigma \phi^\Sigma + \divS (\phi^\Sigma  \, {\bf v}^\Sigma +  \mathbf{j}^\Sigma)
+ [\![ \phi ({\bf v} - {\bf v}^\Sigma ) + \mathbf{j} ]\!]  \cdot \mathbf{n}_{\Sigma}=f^\Sigma \;\; \mbox{ on } \Sigma.
\end{equation}
Note that the velocities ${\bf v}, {\bf v}^\Sigma$ appearing in the convective flux contributions are not yet specified and can also refer to individual velocities ${\bf v}_i$ of constituent $A_i$.
\section{Sharp-Interface balance equations}
We suppose that the fluid phases are composed of $N$ different constituents (chemical species) $A_1,\ldots ,A_N$.
Typically, all species will be present at least in one bulk phase and on the interface. We do not separately index the
constituents in the different phases; if a species is not present, its mass density simply vanishes.
We account for bulk chemical reactions between the $A_i$ according to
\begin{equation}\label{chem-react-bulks}
\alpha_1^a \, A_1 + \ldots + \alpha_N^a \, A_N
\rightleftharpoons
\beta_1^a \, A_1 + \ldots + \beta_N^a \, A_N
\quad \mbox{ for } a=1,\ldots ,N_R
\end{equation}
with stoichiometric coefficients $\alpha_i^a, \beta_i^a \in \N_0$.
Note that we allow for separate chemical reaction networks in the different bulk phases, i.e.\ in full detail
the stoichiometric coefficients read $\alpha_i^{+,a}, \beta_i^{+,a} \in \N_0$ for $a=1,\ldots ,N_R^+$ to describe
the chemistry in $\Omega^+(t)$ and
$\alpha_i^{-,a}, \beta_i^{-,a} \in \N_0$ for $a=1,\ldots ,N_R^-$ for the chemistry in $\Omega^-(t)$.
Since the aspect of chemical reactions is not the main point, here, we suppress the phase index for better readability
whenever this is possible without loss of understanding.
All reactions are considered with their forward and backward direction. The molar reaction rate, i.e.\ the number of chemical conversions in multiples of the Avogadro number per time and volume, of the $a^{\rm th}$ reaction
is denoted $R_a^f$ for the forward (from left to right in \eqref{chem-react-bulks}) and $R_a^b$ for the backward direction.
The total molar reaction rate of the $a^{\rm th}$ reaction hence is $R_a = R_a^f - R_a^b$
and the mass production rate of constituent $A_i$ due to the $a^{\rm th}$ reaction is 
\begin{equation}
r_i = \sum\limits_{a=1}^{N_R} M_i \nu_i^a R_a
\quad \mbox{ with } \nu_i^a:= \beta_i^a - \alpha_i^a,
\end{equation}
where $M_i$ is the molar mass of $A_i$.
Because mass is conserved in every single reaction, it holds that
\begin{equation}\label{mass-conservation-a}
\sum\limits_{i=1}^N M_i \nu_i^a =0 \quad \mbox{ for all } a=1,\ldots ,N_R.
\end{equation}
Recall that we write $A_i^\Sigma$ for the adsorbed form of constituent $A_i$ in order to distinguish it from the latter,
i.e.\ from $A_i^\pm$ in the adjacent bulk phases.
We then account for interfacial chemical reactions between the $A_i^\Sigma$ according to
\begin{equation}
\alpha_1^{\Sigma, a} \, A_1^\Sigma + \ldots + \alpha_N^{\Sigma, a} \, A_N^\Sigma
\rightleftharpoons
\beta_1^{\Sigma, a} \, A_1^\Sigma + \ldots + \beta_N^{\Sigma, a} \, A_N^\Sigma
\quad \mbox{ for } a=1,\ldots ,N_R^\Sigma.
\end{equation}
In full analogy to the bulk case, we let
\begin{equation}
r_i^\Sigma = \sum\limits_{a=1}^{N_R^\Sigma} M_i \nu_i^{\Sigma, a} R_a^\Sigma
\quad \mbox{ with } \nu_i^{\Sigma, a}:= \beta_i^{\Sigma, a} - \alpha_i^{\Sigma, a}
\end{equation}
to obtain $r_i^\Sigma$ for the interfacial production rate of partial mass of $A_i^\Sigma$ and
have
\begin{equation}\label{mass-conservation-a-int}
\sum\limits_{i=1}^N M_i \nu_i^{\Sigma ,a} =0 \quad \mbox{ for all } a=1,\ldots ,N_R^\Sigma.
\end{equation}

Building on \cite{BD2015} and on the appendix in \cite{Bo-interface-mass}, we derive a full multi-velocity model, in which the constituents' partial mass 
as well as their partial momentum are balanced individually, both for the bulk phases and for the interface.
Starting point for this class-II continuum thermodynamics of such systems therefore are the balance equations for partial and total mass, partial and total momentum as well as total internal energy, both in the bulk phases and on the interface.
We also include the balance equations for partial internal energy since it determines certain structural properties.

We start with the balance of partial mass, where we insert
\begin{align*}
\phi & =\rho_i,\quad\quad
\vbf =\vbf_i,\quad\;\;\;
\mathbf{j} =0,\quad\;\;\;
f =r_i,\\
\phi^\Sigma & =\rho_i^\Sigma,\quad\,
\vbf^\Sigma =\vbf_i^\Sigma,\quad\,
\mathbf{j}^{\Sigma} =0,\quad\,
f^{\Sigma} =r_i^{\Sigma}
\end{align*}
into the local form of the master balance \eqref{master-bal-E2} and \eqref{master-bal-int}, respectively.
This yields the\\[1.5ex]
\noindent
{\bf Partial mass balance}
\begin{align}\label{partial-mass-bulk}
& \pa_t \rho_i + \div (\rho_i \vbf_i)= r_i & \mbox{ in } \Omega \setminus \Sigma,\\
& \pa^\Sigma_t \rho^\Sigma_i + \divS(\rho^\Sigma_i \vbf^\Sigma_i)+ [\![\rho_i(\vbf_i-\vbf^\Sigma_i) \cdot \nbf_\Sigma]\!] = r^\Sigma_i & \mbox{ on } \Sigma.\label{partial-mass-int}
\end{align}
Below, we occasionally skip the domains for the balance equations to save space, since the time derivative determines whether the balance
is a bulk or an interface balance.
In order to maintain a single common interface between the two bulk phases, the individual interface velocities
$\vbf^\Sigma_i$ need to fulfil the {\em consistency requirement} (cf.\ \cite{Bo-2PH-ODE})
\begin{align}\label{visigma-consistency}
\vbf^\Sigma_i \cdot \nbf_\Sigma = V_\Sigma \; \mbox{ for all } i=1, \ldots ,N.
\end{align}
In other words, the surfaces of discontinuity of the individual mass densities move with the same normal speed as $\Sigma$.
But note that the individual interfacial velocities will in general have different tangential components.

The jump term in \eqref{partial-mass-int} represents two one-sided bulk-interface mass exchange terms.
In the physico-chemical sciences, these bulk-interface mass (species) exchange processes are named {\rm sorption},
with {\em adsorption} being the bulk-to-interface mass exchange and {\em desorption} for the reverse direction.
We therefore let
\begin{align}
\dot{m}^{\pm,\Sigma}_i := \rho^\pm_i(\vbf^\pm_i- \vbf^\Sigma_i) \cdot \nbf^\pm
\end{align}
with $\nbf^\pm$ the outer normals to $\Omega^\pm$
denote these one-sided bulk-to-interface exchange rates of partial mass.
Then the {\em sorption formulation} of the partial mass balances reads as
\begin{align}
	& \pa_t \rho_i + \div (\rho_i \vbf_i)= r_i & \mbox{ in } \Omega \setminus \Sigma,\\
    &  \pa^\Sigma_t \rho^\Sigma_i + \divS(\rho^\Sigma_i \vbf^\Sigma_i)
 = \dot{m}^{+,\Sigma}_i + \dot{m}^{-,\Sigma}_i + r^\Sigma_i & \mbox{ on } \Sigma.
\end{align}
As a consequence of \eqref{mass-conservation-a}, \eqref{mass-conservation-a-int}, {\em conservation of total mass} holds true, i.e.\
\begin{align}\label{mass-conservation}
	& \sum\limits_{i=1}^N r_i =0, \qquad \sum\limits_{i=1}^N r^\Sigma_i =0.
\end{align}
Summation of the partial mass balances \eqref{partial-mass-bulk} and \eqref{partial-mass-int}, together with \eqref{mass-conservation},
yields the\\[1ex]
{\bf Balance of total mass}
\begin{align}
& \pa_t \rho + \div(\rho \vbf)=0 & \mbox{ in } \Omega \setminus \Sigma,\\
& \pa^\Sigma_t \rho^\Sigma + \div_\Sigma (\rho^\Sigma \vbf^\Sigma)
- [\![ \rho(\vbf -\vbf^\Sigma) \cdot \nbf_\Sigma]\!] =0 & \mbox{ on } \Sigma,
\end{align}
where\vspace{-0.1in}
\begin{align}
	& \rho:= \sum\limits_{i=1}^N \rho_i, \quad \rho \vbf:= \sum\limits_{i=1}^N \rho_i \vbf_i, \quad \rho^\Sigma:= \sum\limits_{i=1}^N \rho^\Sigma_i, \quad \rho^\Sigma \vbf^\Sigma:= \sum\limits_{i=1}^N \rho^\Sigma_i \vbf^\Sigma_i.\vspace{0.2in}
\end{align}
We continue with the balance of partial momenta, where we insert
\begin{align*}
\phi & =\rho_i \vbf_i,\quad\quad
\vbf =\vbf_i,\quad\;\;\;
\mathbf{j} =-\Sbf_i,\quad\;\;\,
f =\rho_i \bbf_i + {\bf f}_i,\\
\phi^\Sigma & =\rho_i^\Sigma \vbf_i^\Sigma,\quad
\vbf^\Sigma =\vbf_i^\Sigma,\quad
\mathbf{j}^{\Sigma} =- \Sbf^\Sigma_i,\quad
f^{\Sigma} =\rho^\Sigma_i\bbf^\Sigma_i + {\bf f}_i^{\Sigma}
\end{align*}
into the local form of the master balance \eqref{master-bal-E2} and \eqref{master-bal-int}, respectively.
This yields the\\[1.5ex]
{\bf Partial momentum balance}
	\begin{align}\label{partial-mom-bulk}
		& \pa_t(\rho_i \vbf_i) + \div(\rho_i \vbf_i \otimes \vbf_i-\Sbf_i)= \rho_i \bbf_i + {\bf f}_i & \hspace{-0.3in}\mbox{in } \Omega \setminus \Sigma,\\
		&	\pa^\Sigma_t (\rho^\Sigma_i \vbf^\Sigma_i) + \divS(\rho^\Sigma_i \vbf^\Sigma_i \otimes \vbf^\Sigma_i -\Sbf^\Sigma_i)+ [\![\dot{m}_i \vbf_i -\Sbf_i \nbf_\Sigma ]\!]= \rho^\Sigma_i \bbf^\Sigma_i+ \fbf^\Sigma_i  & \mbox{ on } \Sigma\label{partial-mom-int}
\end{align}
with ${\bf f}_i$ and ${\bf f}_i^\Sigma$ the rate of momentum exchange between the species and the abbreviation
	\begin{align}
\dot{m}_i := \rho_i (\vbf_i-\vbf^\Sigma_i) \cdot \nbf_\Sigma = \rho_i (\vbf_i-\vbf^\Sigma) \cdot \nbf_\Sigma,
\end{align}
where the second equality comes from \eqref{visigma-consistency}. 
We assume {\em conservation of total momentum} to be valid, hence
\begin{align}
	& \sum\limits_{i=1}^N \fbf_i = 0, \qquad \sum\limits_{i=1}^N \fbf^\Sigma_i =0.
\end{align}
Employing the abbreviation
\begin{align}
  & \dot{m}:=\sum_{i=1}^N \dot{m}_i = \rho (\vbf-\vbf^\Sigma) \cdot \nbf_\Sigma,
\end{align}
this is applied to the sum of the partial momentum balances \eqref{partial-mom-bulk}, \eqref{partial-mom-int} to obtain the
\\[1.5ex]
{\bf Balance of total momentum}
\begin{align}
	& \pa_t (\rho \vbf) + \div(\rho \vbf \otimes \vbf -\Sbf)= \rho \bbf & \mbox{ in } \Omega \setminus \Sigma,\\
	& \pa^\Sigma_t(\rho^\Sigma \vbf^\Sigma)+ \divS(\rho^\Sigma \vbf^\Sigma \otimes \vbf^\Sigma -\Sbf^\Sigma) + [\![\dot{m} \vbf -\Sbf \nbf_\Sigma ]\!]= \rho^\Sigma \bbf^\Sigma & \mbox{ on } \Sigma,
\end{align}
with the total bulk and interface stresses
\begin{align}
	& \Sbf:=\sum\limits_{i=1}^N (\Sbf_i - \rho_i \ubf_i \otimes \ubf_i), \qquad
	& \Sbf^\Sigma:= \sum\limits_{i=1}^N (\Sbf^\Sigma_i -\rho^\Sigma_i \ubf^\Sigma_i \otimes \ubf^\Sigma_i),
\end{align}
containing the bulk and interface diffusion velocities
\begin{align}
	& \ubf_i:= \vbf_i-\vbf, \qquad  \ubf^\Sigma_i := \vbf^\Sigma_i-\vbf^\Sigma,
\end{align}
and the bulk and interface body force densities
\begin{align}
\rho \bbf:= \sum\limits_{i=1}^N \rho_i \bbf_i,\qquad \rho^\Sigma \bbf^\Sigma:= \sum\limits_{i=1}^N \rho^\Sigma_i \bbf^\Sigma_i.
\end{align}
Below, we employ the stress decompositions
\begin{align}\label{stress-splitting}
	& \Sbf_i = -P_i {\bf I} + \Sbf_i^\circ, \quad
	& \Sbf_i^\Sigma = -P_i^\Sigma {\bf I}_\Sigma + \Sbf_i^{\Sigma, \circ}
\end{align}
with the \emph{mechanical partial pressures} $P_i$ and $P_i^\Sigma$ defined as
\begin{align}
	& P_i := - \frac 1 3 {\rm trace} (\Sbf_i), \;\;
	& P_i^\Sigma := - \frac 1 2 {\rm trace} (\Sbf_i^\Sigma ),
\end{align}
where the interfacial 'identity tensor' ${\bf I}_\Sigma$ denotes the projection onto the local tangential plane, i.e.\
\begin{equation}\label{intidentity}
{\bf I}_\Sigma = {\bf I} - {\bf n}_\Sigma \otimes {\bf n}_\Sigma.
\end{equation}
In equilibrium, the mechanical partial pressures reduce to the \emph{thermodynamical partial pressures} $p_i$ and $p_i^\Sigma$, respectively, i.e.\
\begin{align}\label{partial-thermodyn-pressure-i}
	& P_{i|{\rm equ}} = p_i, \;\;
	& P_{i|{\rm equ}}^\Sigma = p_i^\Sigma,
\end{align}
as the partial thermodynamical pressures are functions solely of the respective thermodynamical state variables.
Let
\begin{align}
	& \pi_i := P_i - p_i, \;\;
	& \pi_i^\Sigma := P_i^\Sigma - p_i^\Sigma
\end{align}
denote the \emph{irreversible part of the bulk and interface partial pressures}, respectively. Then
\begin{align}
	& \Sbf_i^{\rm irr} = -\pi_i {\bf I} + \Sbf_i^\circ, \;\;
	& \Sbf_i^{\Sigma, \rm irr} = -\pi_i^\Sigma {\bf I}_\Sigma + \Sbf_i^{\Sigma, \circ}
\end{align}
are the irreversible stress contributions.

We continue with the balance of energy, where we include partial energy balances since they provide
structural information even if they are not part of the final multi-velocity model. 
For simplicity we do not include external energy supply by radiation.
Insertion of
\begin{align*}
\phi & =\rho_i \big(e_i+ \frac{\vbf_i^2}{2}\big),\quad\quad\;\;\;
\vbf =\vbf_i,\quad\;\,
\mathbf{j} = \qbf_i - \vbf_i \cdot \Sbf_i,\quad\;\;
f =\rho_i \vbf_i \cdot \bbf_i + k_i,\\
\phi^\Sigma & =\rho_i^\Sigma \big(e^\Sigma_i+ \frac{(\vbf^\Sigma_i)^2}{2}\big),\;\;
\vbf^\Sigma =\vbf_i^\Sigma,\;\;
\mathbf{j}^{\Sigma} =\qbf^\Sigma_i - \vbf^\Sigma_i\cdot \Sbf^\Sigma_i,\;\;
f^{\Sigma} =\rho^\Sigma_i \vbf^\Sigma_i \cdot\bbf^\Sigma_i + k^\Sigma_i
\end{align*}
into the local form of the master balance \eqref{master-bal-E2} and \eqref{master-bal-int}, respectively, yields the\\[1.5ex]
%
{\bf Partial energy balance}
\begin{align}
	& \pa_t \big(\rho_i \big(e_i+ \frac{\vbf_i^2}{2}\big)\big)+ \div \big(\rho_i \big(e_i+\frac{\vbf_i^2}{2}\big) \vbf_i
+ \qbf_i - \vbf_i \cdot \Sbf_i \big) = \rho_i \vbf_i \cdot \bbf_i + k_i,\\[1ex]
	& \pa^\Sigma_t \big(\rho^\Sigma_i \big(e^\Sigma_i+ \frac{(\vbf^\Sigma_i)^2}{2}\big)\big) + \div_\Sigma \big(\rho^\Sigma_i \big(e^\Sigma_i + \frac{(\vbf^\Sigma_i)^2}{2}\big) \vbf^\Sigma_i +\qbf^\Sigma_i - \vbf^\Sigma_i\cdot \Sbf^\Sigma_i \big) \\
	& + [\![\rho_i \big(e_i + \frac{\vbf_i^2}{2}\big) (\vbf_i-\vbf^\Sigma_i) \cdot \nbf_\Sigma ]\!] + [\![\qbf_i \cdot \nbf_\Sigma ]\!] - [\![\vbf_i \cdot \Sbf_i \nbf_\Sigma]\!] = \rho^\Sigma_i \vbf^\Sigma_i \cdot\bbf^\Sigma_i + k^\Sigma_i,\;\; \nonumber
\end{align}
where $k_i$ and $k_i^\Sigma$ denote the rate of energy exchange between the different constituents.
We assume {\em conservation of total energy}, hence
\begin{align}
	& \sum\limits_{i=1}^N k_i = 0, \qquad \sum\limits_{i=1}^N k^\Sigma_i =0.
\end{align}
Subtraction of the local balance of partial kinetic energy (as obtained from the local balance of partial mass and momentum) yields
the\\[1ex]
{\bf Partial internal energy balance}
\begin{align}\label{partial-intenergy-bulk}
	& \pa_t(\rho_i e_i) + \div(\rho_i e_i \vbf_i+ \qbf_i)= \na \vbf_i : \Sbf_i -\vbf_i \cdot \fbf_i+ \frac{1}{2} r_i \vbf^2_i+ k_i \quad\;\,\mbox{ in }
\Omega \setminus \Sigma,\\[1ex]
	& \pa^\Sigma_t(\rho^\Sigma_i e^\Sigma_i)+ \divS (\rho^\Sigma_i e^\Sigma_i \vbf^\Sigma_i + \qbf^\Sigma_i) + [\![\dot{m}_i (e_i+ \frac{(\vbf_i-\vbf^\Sigma_i)^2}{2})]\!]+ [\![\qbf_i \cdot \nbf_\Sigma]\!]\label{partial-intenergy-int} \\
	& - [\![(\vbf_i-\vbf^\Sigma_i) \cdot \Sbf_i \nbf_\Sigma]\!]
= \na_\Sigma \vbf^\Sigma_i : \Sbf^\Sigma_i- \vbf^\Sigma_i \cdot \fbf^\Sigma_i+ \frac{1}{2} r^\Sigma_i(\vbf^\Sigma_i)^2 + k^\Sigma_i
\quad\quad\,\mbox{ on } \Sigma. \nonumber
\end{align}
We next consider the full mixture internal energy, where the latter has to be carefully defined because there are different options.
In general, the total internal energy is defined to be the remainder if the total kinetic energy is subtracted from the total energy.
In a class-I context, this means to subtract $\rho \vbf^2 /2$ from the total energy density, while all partial kinetic energy
densities $\rho_i \vbf^2_i /2$ should be subtracted in the class-II context; cf.\ \cite{BD2015}. We
congruously employ the latter concept, i.e.\ we define
\begin{align}\label{intenergydensity}
  \rho e = \sum_{i=1}^N \rho_i e_i , \qquad   \rho^\Sigma e^\Sigma = \sum_{i=1}^N \rho_i^\Sigma e_i^\Sigma
\end{align}
to be the densities of the total internal energy in the bulk and on the interface.
This is applied to the sum of equations \eqref{partial-intenergy-bulk} and \eqref{partial-intenergy-int} to obtain the\\[1ex]
{\bf Balance of total internal energy}
\begin{align}
	& \pa_t(\rho e) + \div (\rho e \vbf + \tilde{\qbf} ) = \sum\limits_{i=1}^N \Sbf_i : \na \vbf_i - \sum\limits_{i=1}^N \ubf_i \cdot \big(\fbf_i - r_i \vbf_i + \frac{r_i}{2} \ubf_i\big) \;\mbox{ in }
\Omega \setminus \Sigma,
\end{align}
\begin{align}
	& \pa^\Sigma_t(\rho^\Sigma e^\Sigma) + \div_\Sigma (\rho^\Sigma e^\Sigma \vbf^\Sigma + \tilde{\qbf}^\Sigma)
+ [\![ \dot{m}\, e  ]\!]  + [\![ \tilde{\qbf} \cdot \nbf_\Sigma ]\!]\label{local-total-intenergy} \\
	&  + [\![  \sum\limits_{i=1}^N \dot{m}_i \big( \frac{(\vbf_i -\vbf^\Sigma_i)^2}{2} - \nbf_\Sigma \cdot \frac{\Sbf_i}{\rho_i} \cdot \nbf_\Sigma \big) ]\!]
- [\![ \sum\limits_{i=1}^N (\vbf_i -\vbf^\Sigma_i)_{||} \cdot (\Sbf_i \nbf_\Sigma)_{||} ]\!]\nonumber \\
    & = \sum\limits_{i=1}^N \Sbf^\Sigma_i : \na_\Sigma \vbf^\Sigma_i- \sum\limits_{i=1}^N \ubf^\Sigma_i \cdot \big(\fbf^\Sigma_i - r^\Sigma_i \big(\vbf^\Sigma_i - \frac{\ubf^\Sigma_i}{2}\big)\big) \hspace{1.25in}\mbox{ on } \Sigma\nonumber
\end{align}
with
\begin{align}
	& \tilde{\qbf}:= \sum\limits_{i=1}^N (\qbf_i + \rho_i e_i \ubf_i),
 \quad \tilde{\qbf}^\Sigma = \sum\limits_{i=1}^N \big(\qbf^\Sigma_i + \rho^\Sigma_i e^\Sigma_i \ubf^\Sigma_i\big).
\end{align}
We will later rewrite the latter balances, employing the final form of the heat fluxes which will only become
clear after the entropy production rates have been formulated in an appropriate form.

All constitutive quantities need to be related to the independent variables
by material-dependent functions, i.e.\ by closure relations.
These functions cannot be arbitrary, but need to satisfy certain principles, namely the principle of material frame indifference
and the second law of thermodynamics, i.e.\ the entropy principle including
the entropy inequality. We focus on the latter and only refer to \cite{M85, BD2015} and the references given there
concerning the principle of material frame indifference and other principles which might be imposed.
\section{Entropy Principle - The Second Law of Thermodynamics}
We employ the entropy principle in a sharpened version, which has been introduced in \cite{BD2015}, here extended from single-phase
to two-phase multicomponent fluid systems.
In order to state the entropy principle in an axiomatic form, we assign to every physical quantity an associated {\it parity},
called {\it positive} (+1) or {\it negative} (-1).
The original concept of parity for a time-dependent quantity characterizes its behavior under time reversal (see \cite{Meixner73}),
but in the continuum setting, the symmetry of the solutions under time reversal is lost, because the partial
differential equations (PDEs)
describe irreversible processes. However, for studying the structure of the (unclosed) balance equations,
the concept of reversibility does not play a role.
This observation allows to adapt and extend the notion of parity to the balance equations of continuum mechanics.
It turns out (see \cite{BD2015} for details) that
the parity of a physical quantity is obtained by assigning the factor -1 if the unit of time, 'second' (s), appears with an uneven power and +1 if it appears with even power. Here we rely on the SI base units m, kg, s, K, mol, A, cd. The unit 'ampere' (A) for the electric current does not occur in the present mixture theory\footnote{cf.\ \cite{Dreyer-Guhlke-M} for an extension to electrothermodynamics}. If the latter unit is involved, a further factor -1 is assigned if the unit ampere appears with uneven power. For instance, the combination A\hspace{1pt}s leads to even parity. For example, the parity of the densities of mass, momentum and internal energy are
\begin{equation} \label{parity-examples}
[\rho]=\frac{\rm kg}{\rm m^3} \to+1,\qquad [\rho {\bf v}]=\frac{\rm kg}{\rm m^2 s}\to -1,\qquad [\rho e]=\frac{\rm kg}{\rm m\, s^2}\to+1.
\end{equation}
Evidently, time derivatives alter the parity, while spatial derivatives keep the parity unchanged.

Any solution of the above system of partial
differential equations, composed of the balance equations \eqref{partial-mass-bulk}, \eqref{partial-mass-int},
\eqref{partial-mom-bulk}, \eqref{partial-mom-int}, \eqref{partial-intenergy-bulk} and \eqref{partial-intenergy-int},
is called a {\it thermodynamic process}. Here, by a solution we
mean functions which satisfy the equations in a local sense, i.e.\ they do not form the solution
of a complete initial boundary value problem.
In particular, the value of a quantity and of its spatial derivative can
thus be chosen independently by adjusting the time derivative appropriately.
With this notion, the {\em second law of thermodynamics} consists
of the following axioms.\\[-1ex]
\begin{enumerate}[leftmargin=0pt]
\item[(I)]
There are entropy/entropy-flux pairs $(\rho s, \Phi)$ and $(\rho^\Sigma s^\Sigma, \Phi^\Sigma)$ as material-dependent quantities,
where $\rho s$, $\rho^\Sigma s^\Sigma$ are objective scalars and $\Phi$, $\Phi^\Sigma$ are objective vectors.\vspace{0.05in}
\item[(II)]
The pairs $(\rho s, \Phi)$, $(\rho^\Sigma s^\Sigma, \Phi^\Sigma)$ satisfy the balance equations
\begin{eqnarray}\label{E69}
\pa_t(\rho s)+ \na \cdot (\rho s \vbf+\Phibf)=\zeta \;\;& \mbox{in } \Omega\setminus \Sigma,\quad\\[1ex]
\label{E70}
\! \pa^\Sigma_t(\rho^\Sigma s^\Sigma) + \na_\Sigma \cdot (\rho^\Sigma s^\Sigma \vbf^\Sigma+ \Phibf^\Sigma)+ [\![\rho s(\vbf-\vbf^\Sigma)+\Phibf]\!] \cdot \nbf_\Sigma=\zeta^\Sigma & \mbox{on } \Sigma,\qquad\;\;
\end{eqnarray}
where $\zeta$ and $\zeta^\Sigma$ are the entropy production rates in the bulk and on the interface, respectively.
The specific entropies $s,\, s^\Sigma$ have the physical dimension ${\rm J}\, {\rm kg}^{-1}\, {\rm K}^{-1}={\rm m}^2
{\rm s}^{-2} {\rm K}^{-1}$,
hence are of positive parity. The entropy fluxes and the entropy productions thus have negative parity.\vspace{0.05in}
\item[(III)]
Any admissible pair of entropy fluxes $\Phi$, $\Phi^\Sigma$ is such that\\[0.5ex]
    (i) the resulting entropy production rates $\zeta$ and $\zeta^\Sigma$ both consist of sums of binary products according to
    \begin{equation}\label{M15}
    \zeta = \sum_m \mathcal{N}_m \mathcal{P}_m, \qquad \zeta^\Sigma = \sum_{m^\Sigma} \mathcal{N}_{m^\Sigma}^\Sigma \mathcal{P}_{m^\Sigma}^\Sigma,
    \end{equation}
    where the $\mathcal{N}_m$, $\mathcal{N}_{m^\Sigma}^\Sigma$ denote objective quantities of negative parity, while $\mathcal{P}_m$, $\mathcal{P}_{m^\Sigma}^\Sigma$ refer to objective quantities of positive parity.\\[0.25ex]
    (ii) Each binary product corresponds to a dissipative mechanism, and the set of these must be fixed in advance. Then,
    \begin{equation}\label{strong-form}
    \mathcal{N}_m \, \mathcal{P}_m \geq 0 \;\mbox{ and }\; \mathcal{N}_{m^\Sigma}^\Sigma \, \mathcal{P}_{m^\Sigma}^\Sigma \geq 0
    \end{equation}
    for all $m$, ${m^\Sigma}$ and for every thermodynamic process.\\[1ex]
\hspace{-0.4in} In addition to these universal axioms, we impose two more specific ones
which refer to the most general constitutive models we are interested in. These are:\vspace{0.1in}
\item[(IV)]
For the class of fluid mixtures under consideration, we restrict the dependence of the entropy according to
\begin{equation}\label{E72}
\rho s=\rho s(\rho e, \rho_1, \dots, \rho_N) \quad\text{ and }\quad \rho^\Sigma s^\Sigma=\rho^\Sigma s^\Sigma(\rho^\Sigma e^\Sigma, \rho^\Sigma_1, \dots, \rho^\Sigma_N),
\end{equation}
where $\rho s$ and $\rho^\Sigma s^\Sigma$ are {\it strictly concave functions} which are strictly increasing in the first argument
and satisfy the principle of material frame indifference.
By means of these functions, we define the
{\it (absolute) temperature} $T,\, T^\Sigma$ and {\it chemical potentials} $\mu_i, \, \mu_i^\Sigma$ according to
\begin{equation}\label{E73}
\frac{1}{T}:= \frac{\pa \rho s}{\pa \rho e}, \quad - \frac{\mu_i}{T}:= \frac{\pa \rho s}{\pa \rho_i} \quad\text{and}\quad \frac{1}{T^\Sigma}:= \frac{\pa \rho^\Sigma s^\Sigma}{\pa \rho^\Sigma e^\Sigma}, \quad - \frac{\mu^\Sigma_i}{T^\Sigma}:= \frac{\pa \rho^\Sigma s^\Sigma}{\pa \rho^\Sigma_i}.
\end{equation}
\vspace{0.05in}
\item[(V)]
The following {\it dissipative mechanisms} occur in the fluid mixtures under consideration:
{\it mass diffusion}, {\it chemical reaction},
{\it viscous flow} (including bulk and shear viscosity) and {\it heat conduction} both in the bulk phases and in the interface
as well as {\it sliding friction}, {\it energy transfer} and {\it ad- and desorption} between the two bulk phases and the interface.
\end{enumerate}
Axiom (IV) implicitly contains an assumption of local thermodynamical equilibrium in order
to formulate relations of the type \eqref{E72}. But note that this does not include a statement
of local thermodynamic equilibrium between the interface and the adjacent bulk locations.

A comprehensive discussion of this form of the entropy principle can be found in \cite{BD2015}, \cite{Bo-interface-mass} and the references given there.\\[1ex]
\noindent
In order to utilize the entropy principle, we need to calculate the structure of the entropy production rates
by means of the left-hand sides of the bulk and interface entropy balance equations.
For this purpose, we insert the entropy representations from \eqref{E72} into the entropy balances (\ref{E69}) and (\ref{E70}).
Then, using the chain rule and \eqref{E73}, the balance equations for mass, momentum and internal energy are employed to eliminate
all time derivatives. At this point, we make a general simplification by assuming the fluids to be non-polar in the sense
that the stresses are symmetric, i.e.\
\begin{equation}\label{sym-stress}
{\bf S}_i^{\sf T} = {\bf S}_i   \quad \mbox{ and } \quad \big({\bf S}_i^\Sigma \big)^{\sf T} = {\bf S}_i^\Sigma
\quad \mbox{ for } i=1,\ldots ,N.
\end{equation}
Recall that we split the stresses into isotropic and traceless parts according to \eqref{stress-splitting}.\\

We first consider the bulk entropy production.
We insert $\rho s = \rho s (\rho e,\rho_1, \ldots ,\rho_N)$
into the first two terms  and carry out the differentiations. Then
we eliminate the resulting time derivatives by means of the balance equations \eqref{partial-mass-bulk} and \eqref{local-total-intenergy}, introducing temperature and chemical potentials (defined as in \eqref{E73})
via the chain rule.
After straightforward computations, a first representation of the bulk entropy production is
\begin{align}
\label{entropy-production1}
\zeta =
\div \Big( \Phi - \frac{\tilde{\bf q}}{T} + \sum_{i=1}^N \frac{\rho_i \mu_i {\bf u}_i}{T} \Big)
- \frac 1 T \Big(  \rho e - T \rho s - \sum_{i=1}^N \rho_i \mu_i  \Big)\, \div {\bf v}  \nonumber\\
- \frac 1 T \sum_{i=1}^N  P_i \, \div {\bf v}_i
+ \frac 1 T \sum_{i=1}^N \stress_i^\circ : {\bf D}_i^\circ
- \frac 1 T \sum_{a=1}^{N_R} R_a \mathcal{A}_a
+ \tilde{\bf q} \cdot \nabla \frac 1 T\\
- \sum_{i=1}^N {\bf u}_i \cdot \Big( \rho_i \nabla \frac{\mu_i}{T}
+ \frac 1 T \big( {\bf f}_i - r_i {\bf v}_i  + \frac{1}{2} r_i{\bf u}_i \big) \Big). \nonumber
\end{align}
In \eqref{entropy-production1}, the term $\sum_{i=1}^N r_i \mu_i$ has been rewritten as
$\sum_{a=1}^{N_R} R_a \mathcal{A}_a$, introducing the so-called {\it affinities}
\begin{equation}
\label{affinities}
\mathcal{ A}_a = \sum_{i=1}^N \nu_i^a M_i \mu_i.
\end{equation}
This reflects the fact that each chemical reaction relates to a dissipative mechanism.

Now, for the evaluation of the entropy production in the class-II model one has to account for
the fact that, while for instance $\div {\bf u}_i$ can be chosen independently of ${\bf u}_i$ in exploiting the entropy principle, the defining relations
${\bf u}_i={\bf v}_i-{\bf v}$ lead to the constraints $\div {\bf u}_i=\div {\bf v}_i-\div {\bf v}$.
These constraints are incorporated into the exploitation using Lagrange multipliers $\Lambda_i =\Lambda_i (T, \rho_1, \ldots ,\rho_N)$ and adding the terms
\[
\sum_{i=1}^N \frac{\Lambda_i}{T} (\div {\bf v}_i - \div {\bf v} - \div {\bf u}_i)
\]
to the entropy production. Indeed, by Lemma~7.3.1 in \cite{Liu},
the entropy inequality with this addition and without constraints is equivalent to the original inequality, but evaluated under the constraints.
Then, after rearrangement of terms, the entropy production reads as
\begin{align}
\zeta =
\div \Big( \Phi - \frac{\tilde{\bf q}}{T} + \sum_{i=1}^N \frac{(\rho_i \mu_i -\Lambda_i) {\bf u}_i}{T} \Big)
- \frac 1 T \sum_{a=1}^{N_R} R_a \mathcal{A}_a +
\frac 1 T \sum_{i=1}^N \stress_i^\circ : {\bf D}_i^\circ
\nonumber \\
- \frac 1 T \sum_{i=1}^N \big( P_i - \Lambda_i \big)\, \div {\bf v}_i
- \frac 1 T \Big(  \rho \psi - \sum_{i=1}^N (\rho_i \mu_i -\Lambda_i)  \Big)\, \div {\bf v}
 \label{entropy-inequality-II} \\
+ \tilde{\bf q} \cdot \nabla \frac 1 T
- \sum_{i=1}^N {\bf u}_i \cdot \Big( \rho_i \nabla \frac{\mu_i}{T}- \nabla \frac{\Lambda_i}{T}
+ \frac 1 T \big( {\bf f}_i - r_i {\bf v}_i  + \frac{1}{2} r_i{\bf u}_i \big) \Big), \nonumber
\end{align}
%
where we introduced the \emph{free energy} $\rho \psi = \rho e - \rho s T$.
In order to satisfy axiom~(III-i) in the simplest possible manner, we choose the entropy flux as
\begin{equation}\label{entropyfluxIIa}
\Phi = \frac{\tilde{\bf q}}{T} - \sum_{i=1}^N \frac{(\rho_i \mu_i -\Lambda_i) {\bf u}_i }{T},
\end{equation}
where the $\Lambda_i$ will be determined next.
For this purpose, we make use of
axiom~(III-ii), which requires $\mathcal{N}_m \, \mathcal{P}_m \geq 0$ for each dissipative mechanism, individually.
Therefore, we can first restrict to thermodynamic processes that run with
\[
R_a = 0, \quad \nabla T =0,\quad \mathbf{u}_i = 0,\quad \mathbf{D}_i^\circ =0
\]
and still obtain information about the remaining binary products.
With ${\rm tr}\, ({\bf D}_i )=\div {\bf v}_i$, the entropy inequality implies
\[
\quad T \zeta =
- \sum_{i=1}^N ( P_i - \Lambda_i )\, \div {\bf v}_i
- \big(\rho \psi + \sum_{i=1}^N (\Lambda_i - \rho_i \mu_i) \big) \, \div {\bf v} \geq 0
\]
for this restricted class of processes,
where $\div {\bf v}$ and all $\div {\bf v}_i$ can be varied independently in the exploitation.
For $\div {\bf v}=\div {\bf v}_i =0$ the processes are in equilibrium, where the entropy production assumes a minimum. 
Independent of any closure relation to be chosen later, this implies
\begin{equation}
\label{GDi}
\rho \psi + \sum_{i=1}^N (\Lambda_i - \rho_i \mu_i) = 0,
\end{equation}
as the left-hand side is a function solely of the thermodynamic state, i.e.\ a function of $(T,\rho_1, \ldots, \rho_N)$.
Hence 
\begin{equation}
\label{ep-volvar}
- \sum_{i=1}^N ( P_i - \Lambda_i )\, \div {\bf v}_i \geq 0,
\end{equation}
where the binary product represents the rate of entropy production due to local, species-specific volume variations within the mixture.
This shows that $\Lambda_i$ necessarily equals the partial thermodynamical pressure of constituent $A_i$
introduced in \eqref{partial-thermodyn-pressure-i}, i.e.\
\begin{equation}\label{def-p_i}
\Lambda_i=P_{i|{\rm equ}}=p_i.
\end{equation}
Consequently, $\sum_{i=1}^{N} \Lambda_i =\sum_{i=1}^{N} p_i =p$ is the thermodynamical pressure of the mixture, and 
\eqref{GDi} becomes the {\it Gibbs-Duhem equation}, viz.\
\begin{equation}
\label{Gibbs-Duhem}
\rho \psi + p - \sum_{i=1}^N \rho_i \mu_i = 0.
\end{equation}
Equation \eqref{Gibbs-Duhem} also holds for general thermodynamic processes,
since it  only involves equilibrium quantities. 
It is sometimes also called \emph{Euler} relation.
Note that only a {\it single} Gibbs-Duhem equation results, although individual stress contributions are accounted for.

For deriving a closed PDE model with simple closure relations, we shall be guided by the fundamental relevance of the entropy derivatives, i.e.\ the terms $1/T$, $1/T^\Sigma$ and $-\mu_i/T$, $-\mu_i^\Sigma / T^\Sigma$, as the amplification factors, which
connect changes in the distribution of the conserved quantities with the rate of entropy production.
Knowing $\Lambda_i=p_i$, we therefore split the term $\nabla p_i/T$ in the entropy production and distribute it accordingly.
This yields the final form of the reduced\footnote{Reduced entropy production refers to the outcome after the entropy flux has been chosen.} entropy production rate, which reads as
\begin{align}
 \zeta = &
\; \frac 1 T \sum_{i=1}^N \stress_i^\circ : {\bf D}_i^\circ
- \frac 1 T \sum_{i=1}^N \pi_i \, \div {\bf v}_i
- \frac 1 T \sum_{a=1}^{N_R} R_a \mathcal{A}_a 
+ \sum_{i=1}^N {\bf q}_i \cdot \nabla \frac 1 T\label{entropy-production-bulk-final} \\
& 
- \sum_{i=1}^N {\bf u}_i \cdot \Big( \rho_i \nabla
 \frac{\mu_i}{T} + \frac 1 T ({\bf f}_i - r_i {\bf v}_i
 + \frac{1}{2} r_i {\bf u}_i -\nabla p_i) - (\rho_i  e_i + p_i) \nabla \frac 1 T \Big).\nonumber
\end{align}
Inserting $\Lambda_i=p_i$ into \eqref{entropyfluxIIa} yields
\begin{equation}\label{entropyfluxIIb}
\Phi = \sum_{i=1}^N \Big( \frac{{\bf q}_i}{T}+ \frac{(\rho_i e_i +p_i - \rho_i \mu_i ) {\bf u}_i}{T} \Big).
\end{equation}
As a side remark, observe that even with a single entropy balance, one might now define a partial entropy via
\begin{equation}\label{partial-entropy}
\rho_i s_i := \frac{\rho_i e_i +p_i - \rho_i \mu_i}{T}.
\end{equation}
Then $\rho s = \sum_{i=1}^{N} \rho_i s_i$ by \eqref{Gibbs-Duhem}, hence the total entropy flux attains the form
\begin{equation}\label{total-entropyfluxII}
\rho s {\bf v} + \Phi = \sum_{i=1}^N \Big( \frac{{\bf q}_i}{T} + \rho_i s_i {\bf v}_i \Big).
\end{equation}
One might then introduce individual entropy balances according to
\begin{equation}\label{partial-entropy-balance}
\pa_t(\rho_i s_i)+ \na \cdot (\rho_i s_i \vbf_i + \frac{{\bf q}_i}{T})=\zeta_i,
\end{equation}
in which case the second law of thermodynamics \eqref{E69} attains the form
\begin{equation}\label{2ndlaw-from-partial}
\sum_{i=1}^{N} \zeta_i \geq 0 \quad \mbox{ for any thermodynamic process.}
\end{equation}
Let us note in passing that we can achieve consistency with the classical class-I entropy flux, given as
\begin{equation}\label{entropyfluxII}
\Phi = \frac{\bf q}{T} - \sum_{i=1}^N \frac{\rho_i \mu_i  {\bf u}_i}{T},
\end{equation}
if we define the {\it heat flux} for the multi-temperature class-II model as 
\begin{equation}\label{def:reduced-heat-flux}
{\bf q}:= \sum_{i=1}^N  \big({\bf q}_i + (\rho_i  e_i + p_i) {\bf u}_i \big).
\end{equation}
Defining the partial enthalpy densities according to
\begin{equation}\label{def:partial-enthalpy}
\rho_i h_i := \rho_i  e_i + p_i,
\end{equation}
this heat flux contains the diffusive transport of the partial enthalpies. This allows for an intuitive interpretation as a fluid element carries energy composed of a kinetic part and an enthalpic part.
Employing the heat flux from \eqref{def:reduced-heat-flux} leads to an alternative form of the total internal energy balance, which for the bulk phases reads as
\begin{equation}\label{eq:energy-rewritten}
\pa_t(\rho e) + \div (\rho e \vbf +\qbf) = \sum\limits_{i=1}^N \Sbf^{\rm irr}_i : \na \vbf_i - p \,\div \vbf - \sum\limits_{i=1}^N \ubf_i \cdot \big(\fbf_i -r_i \vbf_i + \frac{r_i}{2} \ubf_i -\na p_i \big).
\end{equation}

Similarly, a lengthy calculation  along the same line of arguments yields the reduced interfacial entropy production rate as
\begin{align}
	\zeta^\Sigma \! = \,
& \frac{1}{T^\Sigma} \sum\limits_{i=1}^N \Sbf^{\Sigma, \circ}_i : {\bf D}^{\Sigma, \circ}_i
- \frac{1}{T^\Sigma} \sum\limits_{i=1}^N \pi_i^\Sigma \nabla_\Sigma \cdot \vbf_i^\Sigma
- \frac{1}{T^\Sigma} \sum\limits_a R^\Sigma_a \mathcal{A}^\Sigma_a
+ \qbf^\Sigma \cdot \na_\Sigma \frac{1}{T^\Sigma}\nonumber \\
	& - \sum\limits_{i=1}^N \ubf^\Sigma_i \cdot \Big(\rho^\Sigma_i \na_\Sigma \frac{\mu^\Sigma_i}{T^\Sigma} + \frac{1}{T^\Sigma} \big(\fbf^\Sigma_i -r^\Sigma_i (\vbf^\Sigma_i-\frac{\ubf^\Sigma_i}{2})- \na_\Sigma p^\Sigma_i \big)\Big)\nonumber \\
	& - \frac{1}{T^\Sigma} \sum\limits_{i=1}^N (\vbf_i^+ - \vbf^\Sigma_i)_{||} \cdot (\Sbf^{+,\rm irr}_i \nbf^+)_{||}
- \frac{1}{T^\Sigma} \sum\limits_{i=1}^N (\vbf_i^- - \vbf^\Sigma_i)_{||} \cdot (\Sbf^{-,\rm irr}_i \nbf^-)_{||}\label{red-CII-ep-int}\\
& - \Big(\frac{1}{T^+} - \frac{1}{T^\Sigma}\Big) (\dot{m}^{+, \Sigma}\, h^+ + \qbf^+ \cdot \nbf^+ )
- \Big(\frac{1}{T^-} - \frac{1}{T^\Sigma}\Big) (\dot{m}^{-, \Sigma}\, h^- + \qbf^- \cdot \nbf^- ) \nonumber \\
 	& + \sum\limits_{i=1}^N \dot{m}_i^{+, \Sigma} \Big(\frac{\mu_i^+}{T^+} - \frac{\mu^\Sigma_i}{T^\Sigma}
+  \frac{1}{T^\Sigma}\Big( \frac{(\vbf_i^+-\vbf^\Sigma_i)^2}{2} - \nbf^+ \cdot \frac{\Sbf^{+, \rm irr}_i}{\rho_i^+} \cdot \nbf^+ \Big) \Big) \nonumber \\
 	& + \sum\limits_{i=1}^N \dot{m}_i^{-, \Sigma} \Big(\frac{\mu_i^-}{T^-} - \frac{\mu^\Sigma_i}{T^\Sigma}
+  \frac{1}{T^\Sigma}\Big( \frac{(\vbf_i^- -\vbf^\Sigma_i)^2}{2} - \nbf^- \cdot \frac{\Sbf^{-, \rm irr}_i}{\rho_i^-} \cdot \nbf^- \Big) \Big), \nonumber
\end{align}
where we have employed the constitutive relation
\begin{align}
	& \Phibf^\Sigma:= \frac{\qbf^\Sigma}{T^\Sigma}- \sum\limits_{i=1}^N \frac{\rho^\Sigma_i \mu^\Sigma_i \ubf^\Sigma_i}{T^\Sigma}
\end{align}
for the entropy flux. Moreover, the \emph{interfacial heat flux} $\qbf^\Sigma$ is defined as
\begin{align}\label{CII-heatflux-int}
& \qbf^\Sigma= \sum\limits_{i=1}^N \big(\qbf^\Sigma_i +  (\rho^\Sigma_i e^\Sigma_i + p^\Sigma_i ) \ubf^\Sigma_i \big).
\end{align}
The first five terms of the interfacial entropy production are in full analogy to the bulk case, while the remaining once represent dissipative mechanisms due to bulk-interface transmission fluxes. For the latter, we expanded the jump brackets into the one-sided terms according to \eqref{jump-resolved}. This needs to be done in order to decompose the jump terms into two binary products, thus arriving at the postulated structure of the entropy production rate.
The derivation of the reduced interfacial entropy production rate also yields the interfacial Gibbs-Duhem (or, Euler) relation, i.e.\
\begin{align}
\quad \rho^\Sigma e^\Sigma + p^\Sigma - \rho^\Sigma s^\Sigma T^\Sigma = \sum\limits_{i=1}^N \rho^\Sigma_i \mu^\Sigma_i.
\end{align}

We finally record an alternative form of the total internal energy balance for the interface, which
employs the heat fluxes $\qbf^\Sigma$ from \eqref{CII-heatflux-int}.
It reads as
\begin{align}
& \pa^\Sigma_t (\rho^\Sigma e^\Sigma) + \div_\Sigma (\rho^\Sigma e^\Sigma \vbf^\Sigma + \qbf^\Sigma) + [\![ \dot{m} \, h + \qbf \cdot \nbf_\Sigma]\!]\nonumber  \\
&  + [\![ \sum\limits_{i=1}^N \dot{m}_i \big(\frac{(\vbf_i- \vbf^\Sigma_i)^2}{2} - \nbf_\Sigma \cdot \Sbf^{\rm irr}_i \cdot \nbf_\Sigma \big)\Big) ]\!]
- [\![ \sum\limits_{i=1}^N (\vbf_i -\vbf^\Sigma_i)_{||} \cdot (\Sbf^{\rm irr} \nbf_\Sigma)_{||} ]\!]\\
&  = \sum\limits_{i=1}^N \Sbf^{\Sigma,{\rm irr}}_i: \na_\Sigma \vbf^\Sigma_i -p^\Sigma \div_\Sigma\, \vbf^\Sigma
- \sum\limits_{i=1}^N \ubf^\Sigma_i \cdot \Big(\fbf^\Sigma_i- r^\Sigma_i \big(\vbf^\Sigma_i- \frac{\ubf^\Sigma_i}{2}\big) -\na_\Sigma\, p^\Sigma_i \Big).\nonumber
\end{align}
\noindent
The remaining step is the modeling of the constitutive quantities via material-depend functions, where the entropy principle
gives restrictions to the constitutive relations for these dissipative mechanisms.
Recall that the axiom~(III-ii) requires all individual contributions to be non-negative.
\section{Closure Relations}
We group the set of constitutive quantities into 'bulk constitutive quantities', 'interface constitutive quantities' and 'transmission constitutive quantities'.
We start with\\[2ex]
\noindent
{\bf Closure relations for bulk constitutive quantities}.\\[1ex]
The first two terms in the bulk entropy production rate \eqref{entropy-production-bulk-final}  refer to
\emph{viscous stresses}, where we employ a linear (in the co-factors) closure. The traceless (deviatoric) part of the partial stress is thus modeled as
\begin{equation}\label{closure-bulk-stress1}
 \stress_i^\circ = 2 \sum_{k=1}^{N} \eta_{ik} {\bf D}_k^\circ,
\end{equation}
while the irreversible contribution to the partial pressure is modeled as
\begin{equation}\label{closure-bulk-stress2}
\pi_i  = - \sum_{k=1}^{N} \lambda_{ik} \div {\bf v}_k.
\end{equation}
The matrices $[\eta_{ik}]$ and $[\lambda_{ik}]$ are symmetric and positive (semi-)definite.
All phenomenological coefficients depend on the basic thermodynamic variables which are, after a Legendre transformation, $T,\rho_1, \ldots ,\rho_N$ inside the bulk phases.
The deviatoric part $\stress_i^\circ$ models the partial stress due to shear, where the $\eta_{ik}$ are the partial shear viscosities. The $\pi_i$ model viscous pressure contributions due to volume variations, where the material-dependent parameters $\lambda_{ik}$ are the partial bulk (or, volume) viscosities.

Next, we recall the closure of the mass production rates according to \cite{BD2015}, i.e.\ the modeling of the \emph{chemical reaction rates}.
This needs to guarantee non-negativity of the corresponding contribution in the entropy production rate, i.e.\ it must ensure that
\begin{equation}\label{chemreact-dissipation}
\zeta_R:=- \sum_{a=1}^{N_R} R_a \frac{\mathcal{A}_a}{RT} =: - \langle {\bf R},  {\bf A} \rangle \geq 0
\quad \mbox{ with } R_a=R_a^f - R_a^b,
\end{equation}
where it is convenient to divide the affinities by $RT$ (with $R$ the universal gas constant) in order to obtain dimensionless (molar-based) chemical potentials.
Since chemical reactions are activated processes which often run far away from equilibrium,
a linear (in the chemical affinities) closure for $R_a$ is not appropriate.
Instead, we exploit the structure $R_a=R_a^f -R_a^b$, where the unidirectional rates satisfy $R_a^f, R_a^b >0$.
Note that $R_a^f, R_a^b \geq 0$ holds by definition, but also the strict inequalities are valid. The latter follows from the fact that $\rho_i$
cannot vanish locally, unless the constituent $A_i$ becomes extinct. But as we include the forward and backward direction for each chemical reaction, and also account for non-degenerate mass diffusion, this cannot happen.
Therefore, we can employ the nonlinear closure\footnote{This form of the closure can be traced back to work by Marcelin
\cite{marcelin1910}, Kohnstamm and, later, by De Donder; cf.\ \cite{lengyel}. I am grateful to V.\ Giovangigli for pointing out the reference \cite{marcelin1910}.}
\begin{equation}
\label{closure-chem-rates}
\ln \frac{R_a^f}{R_a^b}= - \gamma_a \frac{\mathcal{ A}_a}{RT}
\quad \mbox{ with } \gamma_a >0.
\end{equation}
This closure implies a non-negative contribution to the entropy production by chemical reactions in the bulk phases,
since the logarithm is monotone increasing. 
Because of the \emph{strict} monotonicity of the logarithm, the reactive contribution to the entropy
production only vanishes if {\it all reaction are separately in
equilibrium}, i.e.\ all forward and corresponding backward rates
coincide. This is an instance of the {\it principle of detailed
balance}, called Wegscheider's condition in the context of
chemical reaction kinetics.
Note that one of the rates -- either for the forward or the backward path -- still needs to be
modeled, while the form of the other one then follows from \eqref{closure-chem-rates}.

The standard closure for chemical reaction rates, used in so-called {\it elementary reactions} that do not involve intermediate steps,
is referred to as {\it mass action kinetics}.
There, based on the idea of 'reactive collisions', the rate for the forward reaction, say, is modeled as
$R_a^f = k_a^f  \prod_i x_i^{\alpha_i^a}$ with rate constants $k_a ^f >0$; recall that $x_i=c_i/c$ denotes the molar fractions.
To obtain the backward rate via \eqref{closure-chem-rates}, the chemical potentials need to be modeled by material functions
since they appear in the affinities.
For the important class of so-called {\it ideal mixtures}, the chemical potentials obey the relations
\begin{equation}
\label{ideal-mixture}
\mu_i (T,p,x_k) = g_i (T,p) + \frac{RT}{M_i} \log x_i \quad \mbox{ for } i=1,\ldots , N,
\end{equation}
where $g_i (T,p)$ is the Gibbs free energy of the pure component $A_i$ under the temperature and pressure of the mixture. 
The variable $x_k$ inside $\mu_i$ in \eqref{ideal-mixture} stands for the full composition $(x_1, \ldots, x_N)$,
but constrained to satisfy $\sum_k x_k =1$. As only the single molar fraction $x_i$ appears on the right-hand side, this constraint is not effective.
The class of mixture defined by \eqref{ideal-mixture} is motivated by (ideal) mixtures of ideal gases, where
$g_i$ is explicitly given as a function of $T,p$. 
For ideal mixtures, it follows that
\begin{equation}\label{backward-rate}
R_a^b = R_a^f \exp \big( \frac 1 {RT} \sum_{i=1}^{N}g_i (T,p)\big)  \prod_i x_i^{\nu_i^a}.
\end{equation}
Following the above mentioned idea of 'reactive collisions', the forward rate is 
\begin{equation}\label{forward-rate}
R_a^f = k_a^f  \prod_i x_i^{\alpha_i^a} \quad \mbox{ with }  k_a^f>0.
\end{equation}
This leads to a net reaction rate according to
\begin{equation}\label{full-rate}
R_a =  k_a^f  \prod_i x_i^{\alpha_i^a} - k_a^b \prod_i x_i^{\beta_i^a}
\quad \mbox{ with } \frac{k_a^b}{k_a^f}= \exp  ( \frac 1 {RT} \sum_{i=1}^{N}g_i (T,p)),
\end{equation}
showing also the consistency of the present framework in that 'forward' and 'backward' directions can be exchanged.
Note also that the 'rate constants' $k_a^f, k_a^b >0$ will depend on the primitive variables, in fact they depend strongly on the temperature.
It is important to observe that only one of $k_a^f$ or $k_a^b$ can be modeled in this manner,
while the other one is then determined by \eqref{full-rate}.

The result \eqref{full-rate} is restricted to the class of ideal mixtures.
In the case of non-ideal mixtures, the literature usually starts from \eqref{full-rate} as well, but with activities instead of molar
fractions; for a rather exhaustive review see \cite{Pekar}; cf.\ also \cite{lengyel}.
Note that the chemical potentials are then rewritten in the form
\begin{equation}\label{chem-pot-activities}
\mu_i (T,p,x_k) = g_i (T,p) + \frac{RT}{M_i} \log a_i \quad \mbox{ for } i=1,\ldots , N.
\end{equation}
As the so-called activities $a_i$ are allowed to depend on all primitive variables,
the introduction of activities just replaces one unknown function by another one. It is nevertheless widely used in chemical engineering.
With this notation, the general (non-ideal) case leads to a reaction rate of the form
\begin{equation}\label{full-rate-activities}
R_a =  k_a^f  \prod_i a_i^{\alpha_i^a} - k_a^b \prod_i a_i^{\beta_i^a}
\quad \mbox{ with } \frac{k_a^b}{k_a^f}= \exp  ( \frac 1 {RT} \sum_{i=1}^{N}g_i (T,p)).
\end{equation}
Thus the same structure as in \eqref{full-rate} results -- one motivation to work with activities.

Cross-effects between different chemical reactions can also be introduced on the basis of the nonlinear closure for the rate functions from \eqref{closure-chem-rates}. A detailed derivation within the framework of the entropy principle in the form as given above was provided in \cite{BD2015}. 
We recall the final result, viz.\
\begin{equation}
\label{cross-chem-CG}
R_a = \sum_{b=1}^{N_R} \alpha_b L_{ab} \Big( 1 - \exp \big( \sum_{c=1}^{N_R} L_{cb} \frac{\mathcal{A}_c}{RT} \big) \Big)
\end{equation}
with $\alpha_b >0$ and non-negative $L_{ab}$ such that the matrix $L:=[L_{ab}]_{a,b=1}^{N_R}$ is symmetric and positive (semi-)definite, where
all coefficients are functions of $(T,\rho_1, \ldots ,\rho_N)$.
Note that
\begin{equation}\label{chemreact-crosseffects-consistent}
\zeta_R = -\langle L^{-1} {\bf R},  L^{\sf T} \!{\bf A} \rangle
= \sum_{b=1}^{N_R} \alpha_b \big( e^{(L^{\sf T} {\bf A})_b} -1\big) (L^{\sf T} \!{\bf A})_b \geq 0
\; \Leftrightarrow \; \sum_{b=1}^{N_R} (L^{\sf T} \!{\bf A})_b^2 \geq 0.
\end{equation}

Concerning \emph{mass diffusion and heat conduction}, we first specialize to the case without chemical reactions, i.e.\ we first restrict to processes such that $R_a =0$ for all $a$.
The relevant contribution in the entropy production rate then simplifies to read
\begin{equation}
\zeta_{\rm TD} = \sum_{i=1}^N {\bf q}_i \cdot \nabla \frac 1 T
- \sum_{i=1}^N {\bf u}_i \cdot \Big( \rho_i \nabla
 \frac{\mu_i}{T} + \frac 1 T ({\bf f}_i -\nabla p_i) - \rho_i  h_i \nabla \frac 1 T \Big).
\end{equation}
Observe that the term $\rho_i  h_i {\bf u}_i  \nabla \frac 1 T$ naturally fits to both binary products. It hence induces a natural cross-coupling between mass diffusion and heat conduction. Therefore, both phenomena should be treated as a single 'super-mechanism' -- the so-called thermo-diffusion.
Consequently, we will account for general cross-coupling between the two effects, which can be introduced by entropy-neutral mixing,
a concept that has been developed in \cite{BD2015}.
For this purpose, we introduce a zero-addition according to
\begin{equation}\label{thermo-diffusion-bulk1}
\zeta_{\rm TD} = \sum_{i=1}^N ({\bf q}_i -d_i^T \ubf_i) \cdot \nabla \frac 1 T   -   \sum_{i=1}^N {\bf u}_i \cdot 
\Big( \rho_i \nabla  \frac{\mu_i}{T} + \frac{{\bf f}_i -\nabla p_i}{T} - ( \rho_i  h_i + d_i^T ) \nabla \frac 1 T \Big).
\end{equation}
In order to apply a standard closure strategy for the second term,
which has to account for the fact that the co-factors are not all independent, we add $\lambda \sum_i y_i \ubf_i  \nabla \frac 1 T =0$ with a Lagrange multiplier $\lambda \in \R$. This yields
\begin{equation}\label{thermo-diffusion-bulk2}
\zeta_{\rm TD} = \sum_{i=1}^N ({\bf q}_i -d_i^T \ubf_i) \cdot \nabla \frac 1 T   -   \sum_{i=1}^N {\bf u}_i \cdot 
\Big( \rho_i \nabla  \frac{\mu_i}{T} + \frac{{\bf f}_i -\nabla p_i}{T} - ( \rho_i  h_i + d_i^T - y_i \lambda) \nabla \frac 1 T \Big).
\end{equation}
We then choose $\lambda$ such that the co-factors of the $\ubf_i$ add up to zero, which leads to
$\lambda  
 = \sum_{i=1}^{N} d_i^T$,
where the Gibbs-Duhem relation \eqref{Gibbs-Duhem} has been used. Now, replacing the $d_i^T$ by $d_i^T - y_i \lambda$, we see that the introduction of entropy-neutral mixing between the two binary products in $\zeta_{\rm TD}$ leads to \eqref{thermo-diffusion-bulk1}, but with the additional constraint
\begin{equation}\label{td-coefficients-sum}
\sum_{i=1}^{N} d_i^T = 0.
\end{equation}
Now, we are in position to apply the closure in the strict form according to axiom (III), in which every dissipative mechanism is treated separately.
Linear closure yields 
\[
\sum_{i=1}^N ({\bf q}_i -d_i^T \ubf_i) = \alpha \, \nabla \frac 1 T
\]
with a coefficient $\alpha (T,\rho_1,\ldots ,\rho_N)\geq 0$, hence the heat flux from \eqref{def:reduced-heat-flux} ís modeled as
\begin{equation}\label{closure-heat-flux-bulk}
\qbf = \alpha \, \nabla \frac 1 T + \sum_{i=1}^N (\rho_i h_i + d_i^T ) \ubf_i.
\end{equation}
The closure of the momentum exchange terms $\fbf_i$, intimately related to the diffusion velocities $\ubf_i$ and, hence, to mass diffusion, further assumes binary interactions between the species, cf.\ \cite{BD2015} for a detailed derivation. Here, let us only record the result, viz.\
\begin{equation}\label{fi-law-bulk}
\frac 1 T {\bf f}_i = - \rho_i \nabla \frac{\mu_i}{T} + \frac 1 T \nabla p_i +
(\rho_i h_i + d_i^T) \nabla \frac 1 T -  \sum_j f_{ij}  \rho_i \rho_j ( {\bf v}_i  - {\bf v}_j ),
\end{equation}
where the $f_{ij}(T,\rho_1,\ldots ,\rho_N)$ for $i\neq j$ are strictly positive with $f_{ij}=f_{ji}$.

If chemical reactions occur, the $\fbf_i$ are no longer objective vectors, but the $\fbf_i - r_i \vbf_i$ are the constitutive quantities.
For this case, an elaborate derivation has been introduced in \cite{BD2015}, which for the first time gave a consistent treatment of chemical reactions in the class-II framework for single-phase multicomponent fluid systems. The resulting closure law for the interaction forces is
\begin{equation}
\label{bulk-interaction-force-reactive}
 \frac 1 T ( {\bf f}_i - r_i {\bf v}_i )=
- \rho_i \nabla \frac{\mu_i}{T} + \frac 1 T \nabla p_i + (\rho_i h_i+ d_i^T) 
\nabla \frac 1 T - \sum_{j=1}^N (f_{ij} \rho_i \rho_j+C_{ij})
({\bf v}_i - {\bf v}_j)
\end{equation}
with symmetric $f_{ij}(T,\rho_1,\ldots ,\rho_N)>0$, 
where the {\it chemical interaction matrix} has the entries
\begin{equation}
\label{def:Cij}
C_{ij} = \frac 1 T \sum_{a=1}^{N_R} \frac{M_i M_j}{\sum_k \alpha_k^a M_k} \big( R_a^f \beta_i^a \alpha_j^a + R_a^b \alpha_i^a \beta_j^a \big).
\end{equation}
Note that the matrix $[C_{ij}]$ is, in general, {\it not} symmetric.
But it attains a symmetric form if the system is in \emph{detailed chemical reaction equilibrium}, a notion meaning that $R_a^f=R_a^b$
for {\it all} reactions $a=1,\ldots , N_R$. In \cite{BD2015}, besides a motivation of the structure of the $C_{ij}$, it is also shown that this choice is consistent with the entropy principle from above, especially with axiom~(III-ii).

Let us note in passing, that the cross-effects modeled by inclusion of the coupling terms $d_i^T {\bf u}_i  \nabla \frac 1 T$
account for the Soret effect (also called Ludwig-Soret effect), i.e.\ the relative motion of particles due to a temperature gradient,
and the Dufour effect, i.e.\ a diffusive contribution to the heat flux. For a historical review see \cite{RAHMAN2014693}.\\[2ex]
\noindent
{\bf Closure relations for interfacial constitutive quantities}.\\[1ex]
In full analogy to the bulk closure relations we obtain the following constitutive equation.
For the traceless part of the partial interface stresses:
\begin{equation}\label{closure-interface-stress1}
\Sbf^{\Sigma, \circ}_i = 2 \sum_{k=1}^{N} \eta_{ik}^\Sigma {\bf D}^{\Sigma,\circ}_k.
\end{equation}
For the irreversible contribution to the partial interface pressures:
\begin{equation}\label{closure-interface-stress2}
\pi_i^\Sigma  = - \sum_{k=1}^{N} \lambda_{ik}^\Sigma \na_\Sigma \cdot \vbf^\Sigma_k.
\end{equation}
The matrices $[\eta_{ik}^\Sigma]$ and $[\lambda_{ik}^\Sigma]$ are symmetric and positive (semi-)definite.
All phenomenological coefficients depend on the basic thermodynamic variables which are, after a Legendre transformation, $T^\Sigma,\rho_1^\Sigma, \ldots ,\rho_N^\Sigma$.
The $\Sbf^{\Sigma, \circ}_i$ are the partial stresses due to shear inside the interface, where the $\eta_{ik}^\Sigma$ are the partial interface shear viscosities. The $\pi_i^\Sigma$ are the viscous pressure contributions due to area variations inside the interface, where the material-dependent parameters $\lambda_{ik}^\Sigma$ are the partial dilatational viscosities.
The interface shear and dilatational viscosities are also referred to as \emph{intrinsic surface viscosities} in order to distinguish
them from apparent viscosities that can be induced by Marangoni-type effects due to inhomogeneous interfacial tension.
At this point, it should be noted that the interfacial tension, usually denoted as $\gamma$ in the physico-chemical interface sciences and as $\sigma$ in the hydrodynamical theory of capillary flows, is related to the interface pressure via
\[
\gamma \, = \,  \sigma \, = \, - p^\Sigma.
\]
Let us also note in passing that in the physico-chemical interface sciences, the notion of interface (or surface) pressure often refers to
the difference $\gamma^{\rm clean} - \gamma^{\rm contam}$ between the interfacial tension of the clean interface and that of
the interface with, e.g., surface active agents adsorbed to it.

Partial interfacial viscosities have not been studied so far. In the class-I interface fluid model with single (barycentric) velocity, a closure according to \eqref{closure-interface-stress1} and 
\eqref{closure-interface-stress2} gives rise to the so-called \emph{Boussinesq surface fluid}, having linear 'Newtonian' rheology.
The resulting Boussinesq surface stress (\cite{Scriven}, \cite{Slattery-etal-Interfaces}, cf.\ also \cite{BP2010}) reads as
\begin{equation}\label{E20}
\Sbf^\Sigma=(-p^\Sigma+(\lambda^\Sigma-\eta^\Sigma) \na_\Sigma \cdot \vbf^\Sigma)\, {\bf I}_\Sigma+ 2\eta^\Sigma \, \Dbf^\Sigma
\end{equation}
with the interface dilatational and shear viscosities $\lambda^\Sigma, \eta^\Sigma \geq 0$.\\[2ex]
\indent
Closure relations for the chemical reactions rates with cross effects between the different interface reactions:
\begin{equation}
\label{cross-chem-CG-interface}
R_a^\Sigma = \sum_{b=1}^{N_R} \alpha_b^\Sigma L_{ab}^\Sigma \Big( 1 - \exp \big( \sum_{c=1}^{N_R} L_{cb}^\Sigma \frac{\mathcal{A}_c^\Sigma}{RT} \big) \Big)
\end{equation}
with $\alpha_b^\Sigma >0$ and non-negative $L_{ab}^\Sigma$ such that the matrix $L^\Sigma:=[L_{ab}^\Sigma]_{a,b=1}^{N_R^\Sigma}$
is symmetric and positive (semi-)definite, where
all coefficients are functions of $(T^\Sigma,\rho_1^\Sigma, \ldots ,\rho_N^\Sigma)$.\\[1ex]
\indent
Closure relation for the interfacial heat flux:\vspace{-4pt}
\begin{equation}\label{closure-heat-flux-interface}
\qbf^\Sigma = \alpha^\Sigma \, \nabla_\Sigma \frac{1}{T^\Sigma} + \sum_{i=1}^N (\rho_i^\Sigma h_i^\Sigma + d_i^{T,\Sigma} ) \ubf_i^\Sigma,
\end{equation}
with a coefficient $\alpha^\Sigma (T^\Sigma,\rho_1^\Sigma,\ldots ,\rho_N^\Sigma)\geq 0$, 
where the interfacial thermo-diffusion coefficients $d_i^{T,\Sigma}$ satisfy\vspace{-6pt}
\begin{equation}\label{td-coefficients-sum-interface}
\sum_{i=1}^{N} d_i^{T,\Sigma} = 0.
\end{equation}
In \eqref{closure-heat-flux-interface}, $h_i^\Sigma =  e_i^\Sigma + p_i^\Sigma / \rho_i^\Sigma$ denotes the specific partial interfacial enthalpy.\\[1ex]
\indent
Closure relations for the interaction forces between the different constituents, where we focus on the case without chemical reactions for simplicity of presentation:
\begin{equation}\label{fi-law-interface}
\frac{1}{T^\Sigma} {\bf f}_i^\Sigma = - \rho_i^\Sigma \nabla_\Sigma \frac{\mu_i^\Sigma}{T^\Sigma} + \frac{\nabla_\Sigma p_i^\Sigma}{T^\Sigma}  +
(\rho_i^\Sigma h_i^\Sigma + d_i^{T,\Sigma}) \nabla_\Sigma \frac{1}{T^\Sigma}
 -  \sum_j f_{ij}^\Sigma  \rho_i^\Sigma \rho_j^\Sigma ( {\bf v}_i^\Sigma  - {\bf v}_j^\Sigma ),
\end{equation}
where $f_{ij}^\Sigma(T^\Sigma,\rho_1^\Sigma,\ldots ,\rho_N^\Sigma)>0$ (for $i\neq j$) is symmetric.
The case with interfacial chemistry can be treated in full analogy to \eqref{bulk-interaction-force-reactive} and \eqref{def:Cij}.\\[2ex]
\noindent
{\bf Closure relations for bulk-interface transmission constitutive quantities}.\\[1ex]
The remaining binary products in the interfacial entropy production rate from (\ref{red-CII-ep-int}) are associated with the transfer of
partial mass, partial momentum and (internal) energy between one of the bulk phases and the interface.

The $6^{\rm th}$ and $7^{\rm th}$ binary products in (\ref{red-CII-ep-int}) correspond to entropy production due to momentum transfer
between the bulk phases and the interface. This is related to dissipation by friction in case of slippage of (at least one of) the bulk phases against the interface.
Linear (in the co-factors) closure yields individual one-sided Navier slip conditions according to
\begin{equation}\label{E42}
\alpha_i^\pm (\vbf_i^\pm -\vbf_i^\Sigma)_{||} + (\Sbf_i^\pm \cdot \nbf^\pm)_{||}=0 \quad\text{ with }\quad \alpha_i^\pm \geq 0.
\end{equation}
There are two common special choices: the first one is $\alpha_i^\pm=0$, which leads to a free (or, perfect) slip condition, i.e.\ $(\Sbf_i^\pm \cdot \nbf^\pm)_{||}=0$ at $\Sigma$. The second one corresponds to the limit $\alpha_i^\pm \to \infty$, which yields continuity of the tangential velocity at $\Sigma$,
i.e.\ $[\![ \vbf_{i,||} ]\!] =0$, and $\vbf^\Sigma_{i,||} = \vbf_{i,||}$.
In the general case of $\alpha_i^\pm \in (0, \infty )$, the relations \eqref{E42}
link the one-sided limits of the bulk velocity fields to the tangential interface velocities,
where the latter are determined from solving a PDE system, containing a closed version of \eqref{partial-mom-int}.

Let us note in passing that a more general, coupled closure is or course possible.
If cross-effects inside the momentum transmission are accounted for, a corresponding generalisation reads as
\begin{equation}\label{E42cross}
\sum_{k=1}^{N}\alpha_{ik}^\pm (\vbf_k^\pm -\vbf_k^\Sigma)_{||} + (\Sbf_i^\pm \cdot \nbf^\pm)_{||}=0
\qquad (i=1,\ldots ,N),
\end{equation}
where the matrix $[\alpha_{ik}]$ or slip coefficients is symmetric and positive {(se\-mi-)}de\-finite.

The $8^{\rm th}$ and $9^{\rm th}$ binary products in (\ref{red-CII-ep-int}) refer to entropy production due to energy transmission from one of
the bulk phases to the interface.
This is accompanied with a jump in the temperature, i.e.\ the interface temperature $T^\Sigma$ is, in general, different from the one-sided limits of the bulk temperature; see the review paper \cite{Persad-Ward2016review} and the references given there.
Linear closure yields the one-sided closure relations
\begin{equation}\label{E43}
\frac{1}{T^\Sigma}- \frac{1}{T^\pm}=\beta^\pm \big( \dot{m}^{\pm,\Sigma} h^\pm + \qbf^\pm \cdot \nbf^\pm \big) \quad\text{ with }\quad \beta^\pm \geq 0,
\end{equation}
where $h^\pm=e^\pm+ p^\pm / \rho^\pm$ are the specific bulk enthalpies.
A common special choice is $\beta^\pm=0$, implying continuity of the temperature at $\Sigma$,
while $\beta^\pm \to \infty$ leads to an energetically isolating (adiabatic) interface.

The $10^{\rm th}$ and $11^{\rm th}$ binary products in (\ref{red-CII-ep-int}) refer to entropy production due to one-sided mass transfer between the bulk phases and the interface.
If cross-effects between the different species transfer processes are ignored, a linear closure gives the relations
\begin{equation}\label{E44}
\dot{m}^{\pm,\Sigma}_i= \gamma_i^\pm \Big(\frac{\mu^\pm_i}{T^\pm}- \frac{\mu^\Sigma_i}{T^\Sigma}+ \frac{1}{T^\Sigma} \big( \frac{(\vbf_i^\pm-\vbf_i^\Sigma)^2}{2}-\nbf^\pm \cdot \frac{\Sbf_i^{\pm,\rm visc}}{\rho_i^\pm}\cdot \nbf^\pm \big)\Big)
\end{equation}
with $\gamma_i^\pm \geq 0$.
As already noted above in the context of chemical reactions,
one of the rates -- either the ad- or the desorption rate -- has to be modeled based on a micro-theory or experimental knowledge. Then the other rate follows from (\ref{E44}), up to a state-dependent coefficient.
In the limiting cases, as $\gamma_i^\pm \to\infty$, one obtains
\begin{equation}\label{E45}
\frac{\mu^\pm_i}{T^\pm}= \frac{\mu^\Sigma_i}{T^\Sigma}- \frac{1}{T^\Sigma} \big(\frac{(\vbf_i^\pm-\vbf_i^\Sigma)^2}{2}-\nbf^\pm \cdot \frac{\Sbf_i^{\pm,\rm visc}}{\rho_i^\pm}\cdot \nbf^\pm \big).
\end{equation}
These limiting cases correspond to vanishing (one-sided) interfacial resistance against mass transfer. If the temperature is assumed to be continuous, an often imposed assumption, and if the kinetic and viscous terms can be neglected compared to the chemical potentials, then (\ref{E45}) yields
\begin{equation}\label{E46}
\mu^+_i=\mu^-_i=\mu^\Sigma_i;
\end{equation}
cf.\ \cite{1} for an estimation of the strength of the different contributions in (\ref{E45}). 
Let us note that the first identity in (\ref{E46}), which is usually employed to describe the concentration jump at the interface, is often
associated with 'chemical equilibrium' between the bulk phases. This identity has to be evaluated with care, since not only the respective one-sided limits of bulk compositions enter the chemical potentials, but also the different pressures,
so that, in detail, it reads as
\begin{equation}\label{E47}
\mu^+_i(T, p^+, x^+_1, \dots, x^+_N)= \mu^-_i(T, p^-, x^-_1, \dots, x^-_N).
\end{equation}
Since the pressure also has a jump at $\Sigma$ and the height of this jump depends on the curvature, the pressure dependence of the chemical potentials introduces a curvature influence into (\ref{E47}), implying in particular that smaller bubbles display an increased solubility.
This is the Gibbs-Thomson effect, while in the context of thermally driven phase transfer, the analogous effect is described by the Kelvin equation; see \cite{Buff1956, Powles}.

Equation (\ref{E44}) shows that mass transfer is driven by chemical potential differences but, in addition, a kinetic term and viscous forces contribute to the driving force. In mass transfer applications, the systems are usually far from equilibrium in particular
concerning the mass transfer processes. Then, in analogy to the closure of chemical reaction rates,
a non-linear closure turns out to be more appropriate.
Actually, together with the conception of mass transfer as a series of two one-sided
bulk-interface transfer processes, an enriched model for mass transfer results, which includes the local effects of adsorbed surface active agents.
This model was first introduced by the present author in \cite{Bo-interface-mass} and goes far beyond existing mass transfer models.
In the current framework of a multi-velocity model, this will be elaborated on in more detail in the subsequent section.
We refrain from considering cross-couplings between mass transfer of different constituents, the formulation of which would follow along the same lines as already explained above.

Let us finally check that all constitutive quantities as listed above have been treated.
This is clear for all intrinsic bulk and interface quantities, hence we only need to check the bulk-interface transfer terms.
In this regard, we first obtain the $\dot{m}_i^\pm$ from \eqref{E44}, say, or a more refined closure for this term as given below.
Summing up the $\dot{m}_i^\pm$, this also yields $\dot{m}^\pm$. Next, the tangential components of $\Sbf^\pm \nbf_\Sigma$ follow from \eqref{E42}, unless
it is replaced by the condition $ \vbf_{||}^\pm  =\vbf^\Sigma_{||}$ in the limiting case of (one-sided)
no-slip between bulk and interface.
Finally, in the general case of interfacial temperature jumps,
$\qbf^\pm \cdot \nbf_\Sigma$ then follows from \eqref{E43}, since all other quantities therein
are either thermodynamic state variables or belong to the primitive variables, i.e.\ the balanced quantities.

\section{Mass transfer influenced by adsorbed species}
In order to understand the influence of surface coverage by surfactant onto the mass transfer of another chemical species, we have to employ a general closure like (\ref{E44}). However, in mass transfer application, the system is not in local equilibrium at the fluid interface: while the interface is assumed to be in local equilibrium in itself as an interfacial phase, i.e.\ in the tangential directions, the thermodynamic state in adjacent locations in the bulk phases can differ significantly. Hence, in normal direction across the interface, the system can be far away from local equilibrium. Consequently,  a linear closure of mass transfer processes is inappropriate.
We therefore employ a non-linear, logarithmic closure in analogy to chemical reactions.
This leads to constitutive relations for the bulk-interface mass transfer that are consistent to experimentally confirmed closures.
A first version of the derivation below has been given in \cite{Bothe-IBW7}.

We split the one-sided mass transfer terms into an adsorption (superscript 'ad') and a desorption ('de') term, according to
\begin{equation}\label{E16}
\dot{m}^{+,\Sigma}_i=s^{ad,+}_i -s^{de,+}_i, \quad \dot{m}^{-,\Sigma}_i= s^{ad,-}_i -s^{de,-}_i.
\end{equation}
The crucial point is that the transfer
of $A_i$ from bulk phase $+$ to bulk phase $-$, say, is modeled as a series of two sorption processes,
one from phase $+$ onto the interface, followed by another sorption process from $\Sigma$ to bulk phase $-$.
This is consistent with the assumption that the interface, representing a thin transition layer, 
will also carry the transfer component $A_i$
with positive interfacial mass density $\rho_i^\Sigma$, cf.\ Figure~\ref{transfer-series}.
Hence we need to be modeled the one-sided sorption rates, using (non-linear) constitutive relations.

For technical simplicity, we neglect the kinetic and viscous terms inside the mass transfer entropy production contribution, i.e.\ we exploit the reduced binary products
\begin{equation}\label{E48}
\zeta^\pm_{\rm TRANS}= \sum\limits_{i=1}^N \big(s^{ad,\pm}_i- s^{de,\pm}_i\big) \big(\frac{\mu^\pm_i}{T^\pm} - \frac{\mu^\Sigma_i}{T^\Sigma}\big)
\end{equation}
to derive the closure relations, ignoring direct cross-effects between different species. We hence let
\begin{equation}\label{E49}
\ln \frac{s^{ad,\pm}_i}{s^{de,\pm}_i}= a^\pm_i
\big(\frac{\mu^\pm_i}{RT^\pm}-\frac{\mu^\Sigma_i}{R T^\Sigma} \big) \quad\text{ with }\quad a^\pm_i \geq 0.
\end{equation}
As mentioned above in the context of chemical reactions, one of the rates -- either the ad- or the desorption rate -- has to be modeled based on a micro-theory or experimental knowledge. Then the other rate has to obey (\ref{E49}). Below, we let $a^\pm_i=1$ for simplicity as this already suffices to obtain an enriched mass transfer model.

We first consider the case of a soluble surfactant which, for simplicity, is only present in $\Omega^+$, where we suppress the bulk phase index, and on $\Sigma$. 
Desorption is often more easy to model, as there is more space available in the adjacent bulk phase then on the interface.
The simplest realistic rate function is
\begin{equation}\label{E50}
s^{de}_i= k^{de}_i x^\Sigma_i
\end{equation}
with the interfacial molar fraction $x^\Sigma_i=c^\Sigma_i/c^\Sigma$, where $c^\Sigma_i = \rho^\Sigma_i /M_i$ and
$c^\Sigma= \sum_{i=1}^N c^\Sigma_i$.
According to (\ref{E49}) with $a^\pm_i=1$, the associated adsorption rate is
\begin{equation}\label{E51}
s^{ad}_i= k^{de}_i x^\Sigma_i \exp \big(\frac{\mu^\pm_i}{RT^\pm}-\frac{\mu^\Sigma_i}{R T^\Sigma} \big).
\end{equation}
To obtain a concrete model, we assume ideal mixtures both in the bulk and on the interface, i.e.\ (we include all $x_k$
as independent variables, where $\sum_{k=1}^N x_k =1$ is implicitly assumed)
\begin{equation}\label{E52}
\mu^\pm_i(T,p,x_k)= g^\pm_i(T,p)+ RT \ln x^\pm_i
\end{equation}
with $g^\pm_i(T,p)$ denoting the bulk Gibbs free energy of component $A_i$ under the temperature and pressure of the mixture.
Analogously, we employ
\begin{equation}\label{E53}
\mu^\Sigma_i(T^\Sigma,p^\Sigma,x^\Sigma_k)= g^\Sigma_i(T^\Sigma,p^\Sigma)+ RT^\Sigma \ln x^\Sigma_i
\end{equation}
with $g^\Sigma_i(T^\Sigma,p^\Sigma)$ denoting the interface Gibbs free energy of adsorbed component $A_i^\Sigma$ under the interface
temperature and pressure of the mixture. Insertion of the chemical potentials into (\ref{E51}) yields
\begin{equation}\label{E54}
s^{ad}_i= k^{de}_i \exp \big(\frac{g^+_i}{RT} - \frac{g^\Sigma_i}{RT^\Sigma}\big) x^+_i=: k^{ad}_i x^+_i,
\end{equation}
where $k^{ad}_i$ depends in particular on the surface pressure.
Together, \eqref{E50} and \eqref{E54} yield the simplest ad- and desorption rates, leading to the so-called Henry isotherm; see \cite{Kralchevsky} for more details on sorption isotherms of surfactants.

Next, we consider a transfer component $A_i$, like a dissolving gas, which does not accumulate at the interface as a surfactant does, but nevertheless needs to pass through the transmission zone between the bulk phases, mathematically represented by the sharp interface.
Rewriting the interfacial mass balance \eqref{partial-mass-int} for $A_i^\Sigma$ in non-dimensional form, we obtain
\begin{equation}
\label{partial-mass-bal-int-nondim}
\frac{l^{\rm ref}_{\rm n}}{l^{\rm ref}_{\rm tan}} \,
\big( \partial^\Sigma_{t^\ast}  \rho_i^{\Sigma, \ast} 
+ \nabla_\Sigma^\ast \cdot \big( \rho_i^{\Sigma, \ast} {\bf v}_i^{\Sigma, \ast} \big) \big)
 + [ \! [ \dot{m}_i^\ast ] \! ] =0 \mbox{ on } \Sigma,\vspace{-4pt}
\end{equation}
where $l^{\rm ref}_{\rm tan}$ denotes a characteristic length scale along which $\rho_i^\Sigma$ is expected to display significant changes.
The other length scale is $l^{\rm ref}_{\rm n} := \rho_{\rm ref}^\Sigma / \rho_{\rm ref}$.
For non-accumulating species it will characterize
the typical thickness of a layer in the bulk, being adjacent to the interface, such that this layer contains a similar amount of $A_i$ than is adsorbed on the interface (i.e.\ inside the thin transition zone represented by the interface).
So, while $l^{\rm ref}_{\rm tan}$ will represent a mesoscopic length scale, say a bubble or droplet diameter, $l^{\rm ref}_{\rm n}$ will typically be on a sub-micrometer scale.
In this situation, equation \eqref{partial-mass-bal-int-nondim} and, hence, \eqref{partial-mass-int} can be approximated very accurately by
\begin{equation}\label{E55}
[\![\dot{m}_i]\!]=0 \quad \Leftrightarrow \quad \dot{m}^{+,\Sigma}_i+ \dot{m}^{-,\Sigma}_i=0.
\end{equation}
Thus, while the interfacial mixture thermodynamics is accounted for, the capacity of the interface to store species mass is neglected.

Consequently, employing (\ref{E16}) and (\ref{E49}) with $a^\pm_i=1$, we obtain
\begin{equation}\label{E56}
s^{de,+}_i \Big(\exp \big( \frac{\mu^+_i}{RT^+} - \frac{\mu^\Sigma_i}{RT^\Sigma} \big)-1\Big)
+ s^{de,-}_i \Big(\exp \big( \frac{\mu^-_i}{RT^-} - \frac{\mu^\Sigma_i}{RT^\Sigma} \big)-1\Big)=0.
\end{equation}
With the same simplifying assumption of ideal mixtures and (\ref{E50}), this implies
\begin{equation}\label{E57}
k^{de,+}_i \Big(\exp \big( \frac{g^+_i}{RT^+} - \frac{g^\Sigma_i}{RT^\Sigma} \big)x^+_i-x^\Sigma_i\Big)
+ k^{de,-}_i \Big(\exp \big( \frac{g^-_i}{RT^-} - \frac{g^\Sigma_i}{RT^\Sigma} \big)x^-_i-x^\Sigma_i\Big)=0,
\end{equation}
hence
\begin{equation}\label{E58}
\displaystyle
x^\Sigma_i= \frac{k^{de,+}_i \exp \big(\frac{g^+_i}{RT^+} - \frac{g^\Sigma_i}{RT^\Sigma} \big)x^+_i
+ k^{de,-}_i \exp \big( \frac{g^-_i}{RT^-} - \frac{g^\Sigma_i}{RT^\Sigma} \big)x^-_i}{k^{de,+}_i+ k^{de,-}_i}.
\end{equation}
Inserting this value into the first expression in (\ref{E57}), equation (\ref{E16}) yields
\begin{equation}\label{E59}
\dot{m}^{+,\Sigma}_i 
= \frac{k^{de,+}_ik^{de,-}_i}{k^{de,+}_i+ k^{de,-}_i} \exp \big(- \frac{g^\Sigma_i}{RT^\Sigma}\big) 
\Big( \exp \big( \frac{g^+_i}{RT^+}\big) x^+_i- \exp \big( \frac{g^-_i}{RT^-}\big)x^-_i\Big).
\end{equation}
Note that $\dot{m}^{+,\Sigma}_i  = - \dot{m}^{-,\Sigma}_i$ due to \eqref{E55}.
The key point about this closure -- as compared to a closure without interfacial mass densities for transfer components -- is that here, the mass transfer rate is influenced by the interface pressure via the interface Gibbs free energy. This allows to account for the effect of surfactants on the mass transfer of the considered transfer component according to the following causal chain: The presence of surfactant adsorbed at the interface changes the interface pressure. Since the interface chemical potentials of \emph{all} species depend (significantly) on the interface pressure, those change accordingly. 
As the mass transfer rates are not only determined by the one-sided limits of the bulk chemical potentials but by their mutual difference to the interface chemical potential, the transfer rates become dependent on surfactant concentration. This is independent of the hydrodynamics inside the fluid system.

Concrete forms of the mass transfer relation \eqref{E59} of course depend on the employed model for the free energies.
For example, in order to account for the different area demands of the adsorbed species, a possible surface equation of state is
\begin{equation}\label{surfpressb}
p^\Sigma = - K^\Sigma + \frac{RT}{1-\theta}  \sum_{i=1}^{N} \delta_i c^\Sigma_i
\;\; \mbox{ with } \; \theta = \sum_{i=1}^{N} c^\Sigma_i / c_i^{\Sigma , \infty},
\end{equation}
where $\theta$ is the total coverage of the interface and $\delta_i >0$ model the different area demand of $A_i^\Sigma$; cf.\ \cite{RIMS}.
Then the interface analog of the construction of a consistent free energy as explained in
\S 15 in \cite{BD2015} yields the corresponding interface free energy as
\begin{equation}
\rho^\Sigma \psi^\Sigma =  - (1-\theta) p^\Sigma+
RT^\Sigma \sum_{i=1}^{N} \delta_i c^\Sigma_i
\ln \! \big( 1+ \frac{p^\Sigma }{K^\Sigma} \big)
+ RT^\Sigma \sum_{i=1}^N c_i^\Sigma \ln x_i^\Sigma.
\end{equation}
The corresponding (molar based) interfacial chemical potentials then are
\begin{equation}
\mu_i^\Sigma =  g^\Sigma_i(T^\Sigma,p^\Sigma) + RT^\Sigma \ln x_i^\Sigma
\quad \mbox{ for } i=1,\ldots , N
\end{equation}
with\vspace{-0.05in}
\begin{equation}\label{E62b}
g^\Sigma_i(T^\Sigma,p^\Sigma)= \frac{p^\Sigma }{c^{\Sigma,\infty}_i }
+ RT^\Sigma \delta_i \ln \big( 1 + \frac{p^\Sigma}{K^\Sigma} \big) \quad \mbox{ for } i=1,\ldots , N.
\end{equation}
Insertion of (\ref{E62b}) into (\ref{E59}) implies the relation
\begin{equation}\label{mass-transfer-closure}
\dot{m}^{+,\Sigma}_i=  \frac{k_i}{(1 + p^\Sigma /K^\Sigma)^{\delta_i}}
\exp \big( \frac{- p^\Sigma}{c_i^{\Sigma,\infty} RT^\Sigma}\big)
\Big( \exp \big( \frac{g^+_i}{RT^+}\big) x^+_i- \exp \big( \frac{g^-_i}{RT^-}\big)x^-_i\Big).
\end{equation}
Notice that the above derivation treats the adsorbed form $A_i^\Sigma$ of a non-accumulating ('surface inactive') constituent $A_i$
as being an intermediate species in a transition state, when passing through the interface.
It is thus related to transition state theory, often used to derive reaction rates of chemical reactions involving intermediates; cf.\ \cite{laidler}.
Let us also note in passing that in mass transfer applications, a continuous temperature field is usually assumed, i.e.\ $T^+=T^-=T^\Sigma$.

Taking the clean interface as the reference state, and denoting by $\gamma = - p^\Sigma$ the interfacial tension,  \eqref{mass-transfer-closure} implies
\begin{equation}\label{E65b}
\dot{m}^{\rm contam}_i=\Big( \frac{K^\Sigma - \gamma^{\rm clean}}{K^\Sigma - \gamma^{\rm contam}}\Big)^{\delta_i} \,
\exp \Big( - \frac{\gamma^{\rm clean} - \gamma^{\rm contam}}{c_i^{\Sigma,\infty} RT^\Sigma}\Big)\,
\dot{m}^{\rm clean}_i.
\end{equation}
The mass transfer reduction in (\ref{E65b}) contains as the main damping factor the exponential term. This corresponds to an exponential
damping factor of Boltzmann type, i.e.\ a factor
of the form $\exp (-a \, \Delta p^\Sigma /RT^\Sigma)$, in accordance with the energy barrier model due to Langmuir; see \cite{Langmuir}, \cite{Ciani} and the references given there.
An intuitive explanation of the energy barrier model assumes that, in order to cross the interface, a molecule
first has to 'open a whole' in the contaminated interface against the increased surface pressure.
Only a fraction of
the molecules carries sufficient energy in order to perform this work, hence the frequency of transfer of molecules
across the interface is reduced by a corresponding Boltzmann factor.\newpage
%
%
%
%
%
%

\end{document}